\documentclass[prb,aps,eqsecnum,twocolumn,amssymb,showpacs]{revtex4}

\usepackage{color}
\usepackage{graphicx}
\usepackage{longtable}
\usepackage{epsfig}
\usepackage{dcolumn}
\usepackage{bm}
\usepackage{amssymb}
\usepackage{epsfig}
\usepackage{graphicx}
\usepackage{dcolumn}
\usepackage{bm}

\begin{document}

\title{Entangled spin-orbital phases in the bilayer Kugel-Khomskii model }

\author{     Wojciech Brzezicki}
\affiliation{Marian Smoluchowski Institute of Physics, Jagellonian
             University, Reymonta 4, PL-30059 Krak\'ow, Poland }

\author {    Andrzej M. Ole\'{s} }
\affiliation{Max-Planck-Institut f\"ur Festk\"orperforschung,
             Heisenbergstrasse 1, D-70569 Stuttgart, Germany  }
\affiliation{Marian Smoluchowski Institute of Physics, Jagellonian
             University, Reymonta 4, PL-30059 Krak\'ow, Poland }

\date{\today}

\begin{abstract}
We derive the Kugel-Khomskii spin-orbital model for a bilayer and
investigate its phase diagram depending on Hund's exchange $J_H$ and
the $e_g$ orbital splitting $E_z$. In the (classical) mean-field
approach with on-site spin $\langle S_i^z\rangle$
and orbital $\langle \tau_i^z\rangle$
order parameters and factorized spin-and-orbital degrees of freedom,
we demonstrate a competition between the phases with either $G$-type
or $A$-type antiferromagnetic (AF) or ferromagnetic long-range order.
Next we develop a Bethe-Peierls-Weiss method with a Lanczos exact
diagonalization of a cube coupled to its neighbors in $ab$ planes by
the mean-field terms --- this approach captures quantum fluctuations
on the bonds which decide about the nature of disordered phases in the
highly frustrated regime near the orbital degeneracy. We show that the
long-range spin order is unstable in a large part of the phase diagram
which contains then six phases, including also the valence-bond phase
with interlayer spin singlets stabilized by holes in $3z^2-r^2$
orbitals (VB$z$ phase), a disordered plaquette valence-bond (PVB)
phase and a crossover phase between the VB$z$ and the $A$-type AF phase.
When on-site spin-orbital coupling is also included by
the $\langle S_i^z\tau_i^z\rangle$ order parameter,
we discover in addition two entangled spin-disordered phases which
compete with $A$-type AF phase and another crossover phase in between
the $G$-AF phase with occupied $x^2-y^2$ orbitals and the PVB phase.
Thus, the present bilayer model provides an interesting example of
spin-orbital entanglement which generates novel disordered phases.
We analyze the order parameters in all phases and identify
situations where spin-orbital entanglement is crucial and mean-field
factorization of the spin and orbital degrees of freedom leads to
qualitatively incorrect results.
We point out that spin-orbital entanglement may play a role in a
bilayer fluoride K$_3$Cu$_2$F$_7$ which is an experimental realization
of the VB$z$ phase.\\
{\it Published in: Physical Review B \textbf{83}, 214408 (2011).}
\end{abstract}


\maketitle

\section{Introduction}

Recent interest and progress in the theory of spin-orbital
superexchange models was triggered by the observation that orbital
degeneracy drastically increases quantum fluctuations which may
suppress long-range order in the regime of strong competition
between different types of ordered states near the quantum
critical point.\cite{Fei97} The simplest model of this type is the
Kugel-Khomskii ($d^9$) model introduced long ago\cite{Kug82} for
KCuF$_3$, a strongly correlated system with a single hole within
degenerate $e_g$ orbitals at each Cu$^{2+}$ ion. Kugel and
Khomskii showed that many-body effects could then give rise to
orbital order stabilized by a purely electronic superexchange
mechanism. A similar situation occurs in a number of compounds
with active orbital degrees of freedom, where strong on-site
Coulomb interactions localize electrons (or holes) and give rise
to spin-orbital superexchange. \cite{Nag00,Kha05,Hfm} The orbital
superexchange may stabilize the orbital order by itself, but in
$e_g$ systems it is usually helped by the orbital interactions
which follow from the Jahn-Teller distortions of the
lattice.\cite{Kug82,Zaa93,Fei99,Ole05} For instance, in LaMnO$_3$
these contributions are of equal importance and both of them are
necessary to explain the observed high temperature of the
structural transition.\cite{Fei99} Also in KCuF$_3$ the lattice
distortions play an important role and explain its strongly
anisotropic magnetic and optical properties.
\cite{Ole05,Dei08,Leo10}

An important feature of spin-orbital superexchange, which arises
in transition metal oxides with active orbital degrees of freedom,
\cite{Kug82,Nag00,Kha05,Hfm} is generic frustration of the orbital
part of the superexchange. It follows from the directional nature
of orbital interactions,\cite{Fei97} which is in contrast to the
SU(2) symmetry of spin interactions. Therefore, the orbital part
of the spin-orbital superexchange is intrinsically frustrated also
on lattices without geometrical frustration, such as the
three-dimensional (3D)
perovskite lattice of KCuF$_3$ or LaMnO$_3$. Generic features of
this direction-dependent orbital interactions are best captured
within the two-dimensional (2D) quantum compass model,
\cite{Nus04} which exhibits a quantum phase transition from one to
the other one-dimensional (1D) columnar order through a point with
isotropic and strongly frustrated interactions.\cite{Wen08,Oru09}
In spite of the intrinsic frustration and high degeneracy of the
ground state, the long-range order of 1D type exists in the 2D
quantum compass model, as shown by a rigorous proof.\cite{You10}
Numerical simulations demonstrate that this model is in the
universality class of the 2D Ising model\cite{Nus08} and the order
persists in a range of finite temperature.\cite{Wen08} In
contrast, the superexchange interactions for the 2D $e_g$ orbital
model contain orbital quantum fluctuations on the
bonds,\cite{Fei97,vdB99} but nevertheless the long-range order
survives also in this case.\cite{You07}

An intriguing situation arises when spin and orbital part of the
superexchange are strongly coupled and compete with each other, as
found in realistic spin-orbital models for several transition metal
oxides.\cite{Kha05,Hfm} For instance, a qualitatively new spin-orbital
liquid phase may arise when the superexchange interactions are
geometrically frustrated on the triangular lattice,\cite{Nor08}
or spin order cannot stabilize in LiNiO$_2$, another compound with
triangular lattice of magnetic, in spite of presence of strong orbital
interactions which suggest pronounced orbital order.\cite{Ver04}
A more standard situation is found in
the transition metal oxides which crystallize in the perovskite
lattice, where in general spin order coexists with orbital order,
\cite{Nag00,Kha05,Hfm} and both satisfy the classical
Goodenough-Kanamori rules.\cite{Goode} A well known example is the
archetypical compound with degenerate $e_g$ orbitals, KCuF$_3$, in
which the orbital order is stabilized jointly by the superexchange
and Jahn-Teller lattice distortions.\cite{Ole05,Leo10} As a result,
the magnetic interactions are strongly anisotropic and give rise to
quasi-1D Heisenberg antiferromagnetic (AF) chain dominated by
quantum fluctuations and characterized by spinon excitations,
\cite{Ten93} with a dimensional crossover occurring when
temperature is lowered below the N\'eel temperature
$T_N$.\cite{Lak05}

While the coexisting $A$-type AF ($A$-AF) order and the orbital
order is well established in KCuF$_3$ below $T_N$,\cite{Cuc02} and
this phase is reproduced by the spin-orbital $d^9$ superexchange
model,\cite{Ole00} the model poses an interesting question by itself:
Which types of coexisting spin and orbital order (or disorder) are
possible when its microscopic parameters are varied? So far, it was
only established that the long-range AF order is destroyed by strong
quantum fluctuations,\cite{Ole00,Fei98} and it has been shown that
instead certain spin disordered phases with valence-bond (VB)
correlations stabilized by local orbital correlations are favored.
\cite{Fei97,Kha05}
However, the phase diagram of the Kugel-Khomskii $d^9$ model is
unknown --- it was not studied systematically beyond the mean-field
(MF) approximation and certain simple variational wave functions and
it remains an outstanding problem in the theory.\cite{Fei97}

The purpose of this paper is to analyze a simpler situation of the
spin-orbital Kugel-Khomskii model for a bilayer, called below bilayer
spin-orbital $d^9$ model, consisting of two
$ab$ layers connected by interlayer bonds along the $c$ axis. This
choice is motivated by an expected competition of the long-range
AF order with VB-like states. One of them, a VB phase with spin
singlets on the interlayer bonds (VB$z$ phase), is stabilized by large
crystal field $E_z$ which favors occupied $3z^2-r^2$ orbitals (by
holes). We shall investigate the range of stability of this and other
phases, including the $A$-AF phase similar to the one found in KCuF$_3$.

To establish reliable results concerning short-range order in the
crossover regime between phases with long-range AF or FM order, we
developed a {\it cluster MF approach\/} which goes beyond the
single site MF in the spin-orbital system\cite{DeS03} and is based
on an exact diagonalization of an eight-site cubic cluster coupled
to its neighbors by MF terms. This unit is sufficient for
investigating both AF phases with four sublattices and VB states,
with spin singlets either along the $c$ axis or within the $ab$
planes. This theoretical method is motivated by possible
spin-orbital entanglement\cite{Ole06} which is particularly
pronounced in the 1D SU(4) [or SU(2)$\otimes$SU(2)] spin-orbital
models,\cite{Fri99} and occurs also in the models for perovskites
with AF spin correlations on the bonds where it violates the
Goodenough-Kanamori rules. \cite{Goode} In the perovskite
vanadates such entangled states play an important role in their
optical properties,\cite{Kha04} in the phase diagram\cite{Hor08}
and in the dimerization of FM interactions along the $c$ axis in
the $C$-AF phase of YVO$_3$.\cite{Ulr03,Sir08} Below we shall
investigate whether entangled states could play a role in the
present Kugel-Khomskii model for a bilayer with nearly degenerate
$e_g$ orbitals. Thereby we establish exotic type of spin-orbital
order stabilized by joint quantum spin-orbital fluctuations, and
investigate signatures of entangled states in this phase.

The paper is organized as follows. In Sec.~\ref{sec:som} we present the
Kugel-Khomskii $d^9$ spin-orbital model for a bilayer which consists of
two 2D square lattices in $ab$ planes coupled by vertical bonds along
the $c$ axis. First in Sec.
\ref{sec:model} we introduce the $d^9$ spin-orbital model for a
bilayer derived here following Ref. \onlinecite{Ole00}. Its classical
phase diagram obtained in a single-site MF approximation is presented
in Sec. \ref{sec:ssa}. Next we argue that quantum fluctuations and
intrinsic frustration of the superexchange near the orbital degeneracy
motivate the solution of this model in a better MF approximation based
on an embedded cubic cluster, which we introduce in Sec. \ref{sec:mfa}.
It leads to MF equations which were solved self-consistently in an
iterative way, as described in Sec. \ref{sec:ite}. In Sec.
\ref{sec:phd} we present two phase diagrams obtained from the MF
analysis using Bethe-Peierls-Weiss cluster method:
(i) the phase diagram which follows from factorization of spin and
orbital degrees of freedom in Sec. \ref{sec:dis}, and
(ii) the one obtained when also on-site joint on-site spin-orbital order
parameter is introduced, see Sec. \ref{sec:sof}. The latter approach
gives nine different phases, and we describe characteristic features of
their order parameters in Sec. \ref{sec:ops}. We introduce bond
correlation functions in Sec. \ref{sec:nn}, and concentrate their
analysis on the regime of almost degenerate $e_g$ orbitals, focusing on
the proximity of the plaquette VB (PVB) and entangled
spin-orbital (ESO) phases in Secs. \ref{sec:pvb} and \ref{sec:eso}.
Finally, we quantify the spin-orbital entanglement using on-site
and bond correlations, see Sec. \ref{sec:enta}, which modifies
significantly the phase diagram of the model with respect to the
one obtained when spin and orbital operators are disentangled.
General discussion and summary are presented in
Sec.~\ref{sec:summa}.

\section{Spin-orbital model and methods}
\label{sec:som}

\subsection{Kugel-Khomskii model for a bilayer}
\label{sec:model}

For realistic parameters the late transition metal oxides or
fluorides are strongly correlated and electrons localize in the
$3d$ orbitals, \cite{Ole87,Gra92} leading to Cu$^{2+}$ ions with
spin $S=1/2$ in $d^9$ configuration, as e.g. in KCuF$_3$ or
La$_2$CuO$_4$. The virtual charge excitations lead then to
superexchange which involves also orbital degrees of freedom in
systems with partly filled degenerate orbitals. In analogy to the
models introduced for bilayer manganite,\cite{Ole03,Dag06}
La$_{2-x}$Sr$_x$Mn$_2$O$_7$, we consider here a model for
K$_3$Cu$_2$F$_7$ bilayer compound, with two active and nearly
degenerate $e_g$ orbitals,
\begin{eqnarray}
\label{eg} |x\rangle\equiv (x^2-y^2)/\sqrt{2}, \hskip .7cm
|z\rangle\equiv (3z^2-r^2)/\sqrt{6}\,,
\end{eqnarray}
while $t_{2g}$ orbitals do not contribute and are filled with
electrons. They do not couple to $e_g$'s by hopping through fluorine
and hence can be neglected. We investigate in what follows an
electronic model and neglect coupling to the lattice distortions
arising due to Jahn-Teller effect. The bilayer K$_3$Cu$_2$F$_7$ system
is known since twenty years,\cite{Von81} but its magnetic properties
were reported only recently.\cite{Man07} We shall address the orbital
order and magnetic correlations realized in this system below.

The Hamiltonian for $d^9$ systems contains: holes' kinetic energy
$H_{t}$ with hopping amplitude $t$, electron-electron interactions
$H_{\rm int}$, with on-site Hubbard $U$ and Hund's exchange coupling
$J_H$, as well as crystal-field splitting term $H_z$ playing a role
of external orbital field $E_z$ acting on $e_g$ orbitals:
\begin{eqnarray}
H_{e_g}=H_{t}+H_{\rm int}+H_z .
\end{eqnarray}
Because of the shape of the two $e_g$ orbitals Eq. (\ref{eg}), the
effective hopping elements are direction dependent and different
depending on the direction of the bond $\langle ij\rangle$. The only
non-vanishing $(dd\sigma)$ hopping element in the $c$ direction
connects two $|z\rangle$ orbitals,\cite{Zaa93} while the elements in
the $ab$ planes satisfy Slater-Koster relations.

Taking the effective $(dd\sigma)$ hopping element $t$ for two $z$
orbitals on a bond along the $c$ axis as a unit, $H_{t}$ is given by
\begin{eqnarray} \label{Ht}
H_{t}&=&\frac{t}{4}\!\sum_{\langle ij\rangle\parallel ab}
\left\{3d^{\dagger}_{ix\sigma}d^{}_{jx\sigma}
+d^{\dagger}_{ix\sigma}d^{}_{jz\sigma})\right. \nonumber \\
& &\left.\hskip 1.2cm \pm\sqrt{3}(d^{\dagger}_{iz\sigma}d^{}_{jx\sigma}
+d^{\dagger}_{ix\sigma}d^{}_{jz\sigma})+{\rm H.c.}\right\} \nonumber \\
&+&t\sum_{\langle ij\rangle\parallel
c}(d^{\dagger}_{iz\sigma}d^{}_{jz\sigma}+{\rm H.c.}),
\end{eqnarray}
where $d^{\dagger}_{ix\sigma}$ and $d^{\dagger}_{iz\sigma}$ are
creation operators for a hole in $x$ and $z$ orbital with spin
$\sigma=\uparrow,\downarrow$, and the in-plane $x$--$z$ hopping
depends on the phase of $|x\rangle$ orbital involved in the
hopping process along the bond $\langle ij\rangle$ and is included
in the alternating sign of the terms $\propto\sqrt{3}$ between $a$
and $b$ cubic axes. The on-site electron-electron interactions are
described by:\cite{Ole83}
\begin{eqnarray} \label{Hint}
H_{\rm int}&=&U\sum_{i\alpha}n_{i
\alpha\uparrow}n_{i\alpha\downarrow}+(U-3J_H)
\sum_{i\sigma}n_{ix\sigma}n_{iz\sigma} \nonumber \\
&+&(U-2J_H)\sum_{i\sigma}n_{ix\sigma} n_{iz\bar
{\sigma}}-J_H\!\sum_{i\sigma}\!
d^{\dagger}_{ix\sigma}d^{}_{ix\bar{\sigma}}
d^{\dagger}_{iz\bar{\sigma}}d^{}_{iz\sigma} \nonumber \\
&+&J_H\!\sum_i(d^{\dagger}_{ix\uparrow}
d^{\dagger}_{ix\downarrow}d^{}_{iz\downarrow}
d^{}_{iz\uparrow}+H.c.)\,.
\end{eqnarray}
Here $n_{i\alpha\sigma}$ stands for the hole density operator in
orbital $\alpha=x,z$ with spin $\sigma$, and
$\bar{\sigma}=-\sigma$. This Hamiltonian describes the multiplet
structure of $d^8$ or $d^2$ ions and is rotationally invariant in
the orbital space. We assumed the wave function to be real which
gives the same amplitude $J_H$ for Hund's exchange interaction and for
pair hopping term between $|x\rangle$ and $|z\rangle$ orbitals. The
last term of the $H_{e_g}$ Hamiltonian lifts the degeneracy of the two
$e_g$ orbitals
\begin{eqnarray} \label{Hz}
H_{z}=-\frac12 E_z\sum_{i\sigma}(n_{ix\sigma}-n_{iz\sigma}),
\end{eqnarray}
and favors hole occupancy of $x$ ($z$) orbitals when $E_z>0$ ($E_z<0$).
It can be associated with a uniaxial pressure along the $c$ axis,
induced by the bilayer geometry or by external pressure.

The typical energies for the Coulomb $U$ and Hund's exchange $J_H$
elements can be deduced from the atomic spectra or found using
density functional theory with constrained electron densities.
Earlier studies performed within the local density approximation (LDA)
gave rather large values of the interaction parameters:\cite{Gra92}
$U=8.96$ eV and $J_H=1.19$ eV. More recent studies used the LDA with
on-site Coulomb interaction treated within the LDA+$U$ scheme and
gave somewhat reduced values:\cite{Lic95} $U=7.5$ eV and $J_H=0.9$ eV.
However, both parameter sets give rather similar values of Hund's
exchange parameter,
\begin{eqnarray} \label{eta}
\eta=\frac{J_H}{U},
\end{eqnarray}
being close to 0.13 or 0.12, i.e., within the expected range
$0.1<\eta<0.2$ for strongly correlated late transition metal oxides.
Note that the physically acceptable range which follows from
Eq. (\ref{Hint}) is much broader, i.e., $0<\eta<1/3$.

The value of effective intersite $(dd\sigma)$ hopping element $t$ is
more difficult to estimate. It follows from the usual effective process
via the oxygen orbitals described by a $t_{pd}$ hopping, and the energy
difference between the $3d$ and $2p$ orbitals involved in the hopping
process, so-called charge-transfer energy.\cite{Zaa93} A representative
value of $t\simeq 0.65$ eV may be derived from the realistic parameters
\cite{Gra92} of CuO$_2$ planes in La$_2$CuO$_4$. Taking in addition
$U=7.5$ eV, one finds the superexchange constant between hole $S=1/2$
spins within $|x\rangle$ orbitals in a single CuO$_2$ plane,
$J_x=(9/4)t^2/U\simeq 0.127$ eV, which reproduces well the experimental
value, as discussed in Ref. \onlinecite{Ole00}.

Thanks to $t\ll U$ we can safely assume that the ground state is
insulating at the filling of one hole localized at each Cu$^{2+}$
ion. In the atomic limit ($t=0$ and $E_z=0$) we have large
$4^N$-fold degeneracy as the hole can occupy either $x$ or $z$ orbital
and have up or down spin. This high degeneracy is lifted due to
effective superexchange interactions between spins and orbitals at
nearest neighbor Cu ions $i$ and $j$ which act along the bond
$\langle ij\rangle$. They originate from the virtual transitions to the
excited states, i.e., $d^9_id^9_j\rightleftharpoons d^{10}_id^8_j$, and
are generated by the hopping term Eq. (\ref{Ht}). Hence, the effective
spin-orbital model can be derived from the atomic limit
Hamiltonian containing interaction Eq. (\ref{Hint}) and the
crystal-field term Eq. (\ref{Hz}), treating the kinetic term Eq.
(\ref{Ht}) as a perturbation. Taking into account the full
multiplet structure of the excited states for the $d^8$
configuration,\cite{Ole00} one gets the corrections of the order of
$J_H$ to the Hamiltonian which results for the degenerate excited
states (at $J_H=0$). Calculating the energies of the excited $d^8$
states we neglected their dependence on the crystal-field
splitting $E_z$. This assumption is well justified as the
deviation from the equidistant spectrum at $E_z=0$ become
significant only for $|E_z|/J_H>1$ and in case of La$_2$CuO$_4$
one finds $|E_z|/J_H\approx 0.27$. For systems close to orbital
degeneracy, which we are interested in, this ratio is even
smaller.

The derivation which follows Ref. \onlinecite{Ole00} leads to the
spin-orbital model, with the Heisenberg Hamiltonian for the spins
coupled to the orbital problem, as follows:
\begin{eqnarray} \label{hamik}
{\cal H}&=&-\frac{1}{2}J\!\!\sum_{\langle ij\rangle||\gamma}
\left\{(r_1\,\Pi_t^{(ij)}+r_2\,\Pi_s^{(ij)})
\left(\frac{1}{4}-\tau^{\gamma}_i\tau^{\gamma}_j\right)\right. \nonumber \\
& &\hskip .7cm
+\left.\left(r_2+r_4\right)\Pi_s^{(ij)}\left(\frac{1}{2}-\tau^{\gamma}_i\right)
\left(\frac{1}{2}-\tau^{\gamma}_j\right)\right\}  \nonumber \\
&-&E_z\sum_{i}\tau_i^c\,.
\end{eqnarray}
Here $\gamma=a,b,c$ labels the direction of a bond $\langle ij\rangle$
in the bilayer system.
The energy scale is given by the superexchange constant,
\begin{eqnarray}
J=\frac{4t^2}{U}\,,
\end{eqnarray}
and the orbital operators at site $i$ are given by
$\vec\tau_i=\{\tau_i^a,\tau_i^b,\tau_i^c\}$. The terms
proportional to the coefficients $\{r_1,r_2,r_4\}$ refer to the
charge excitations to the upper Hubbard band\cite{Ole00} which
occur in the $d^9_id^9_j\rightleftharpoons d^9_id^{10}_j$
processes and depend on Hund's exchange parameter $\eta$ Eq.
(\ref{eta}) via the coefficients:\cite{noteri}
\begin{eqnarray}
r_1=\frac{1}{1-3\eta},\hskip .7cm r_2=\frac{1}{1-\eta},\hskip .7cm
r_4=\frac{1}{1+\eta}.
\end{eqnarray}
The model Eq. (\ref{hamik}) depends thus on two parameters: (i) Hund's
exchange coupling $\eta$ Eq. (\ref{eta}), and (ii) the crystal-field
splitting $E_z/J$.

The operators $\Pi^s_{ij}$ and $\Pi^t_{ij}$ stand for projections of
spin states on the bond $\langle ij\rangle$ on a singlet ($\Pi^s_{ij}$)
and triplet ($\Pi^t_{ij}$) configuration, respectively,
\begin{eqnarray}
\label{proje}
\Pi_s^{(ij)}=\left(\frac{1}{4}-{\bf S}_i\cdot{\bf S}_j\right),
\hskip .5cm \Pi_t^{(ij)}=\left(\frac{3}{4}+{\bf S}_i\cdot{\bf
S}_j\right),
\end{eqnarray}
for spins $S=1/2$ at both sites $i$ and $j$,
and $\tau^{\gamma}_i$ (with $\gamma=a,b,c$ standing for a direction
in the real space) represent $e_g$ orbital degrees of freedom and
can be expressed in terms of Pauli matrices
$\{\sigma^x_i,\sigma^y_i,\sigma^z_i\}$ in the following way:
\begin{eqnarray}
\tau^{a,b}_i\equiv\frac{1}{4}(-\sigma^z_i\pm\sqrt{3}\sigma^x_i),
\hskip .5cm \tau^c_i\equiv\frac{1}{2}\sigma^z_i.
\end{eqnarray}
The matrices $\{\sigma^{\gamma}_i\}$ act in the orbital space
(and have nothing to do with the physical spin ${\bf S}_i$ present
in this problem). Note
that $\tau^{\gamma}_i$ operators are not independent because they
satisfy the local constraint, $\sum_{\gamma}\tau^{\gamma}_i\equiv 0$.

In Fig. \ref{orbs} we present typical orbitals configurations with
ferro-orbital (FO) order and alternating orbital (AO) order
considered in the $e_g$ orbital models.\cite{Fei97,vdB99} In the
next sections we shall analyze their possible coexistence with
spin order in the bilayer $d^9$ spin-orbital model Eq.
(\ref{hamik}). As we can see, the maximal (minimal) value of the
orbital operators $\tau^{\gamma}_i$ is related with orbital taking
shape of a clover (cigar) with symmetry axis pointing along the
direction $\gamma$.

\subsection{Single-site mean-field approximation}
\label{sec:ssa}

\begin{figure}[t!]
    \includegraphics[width=7.7 cm]{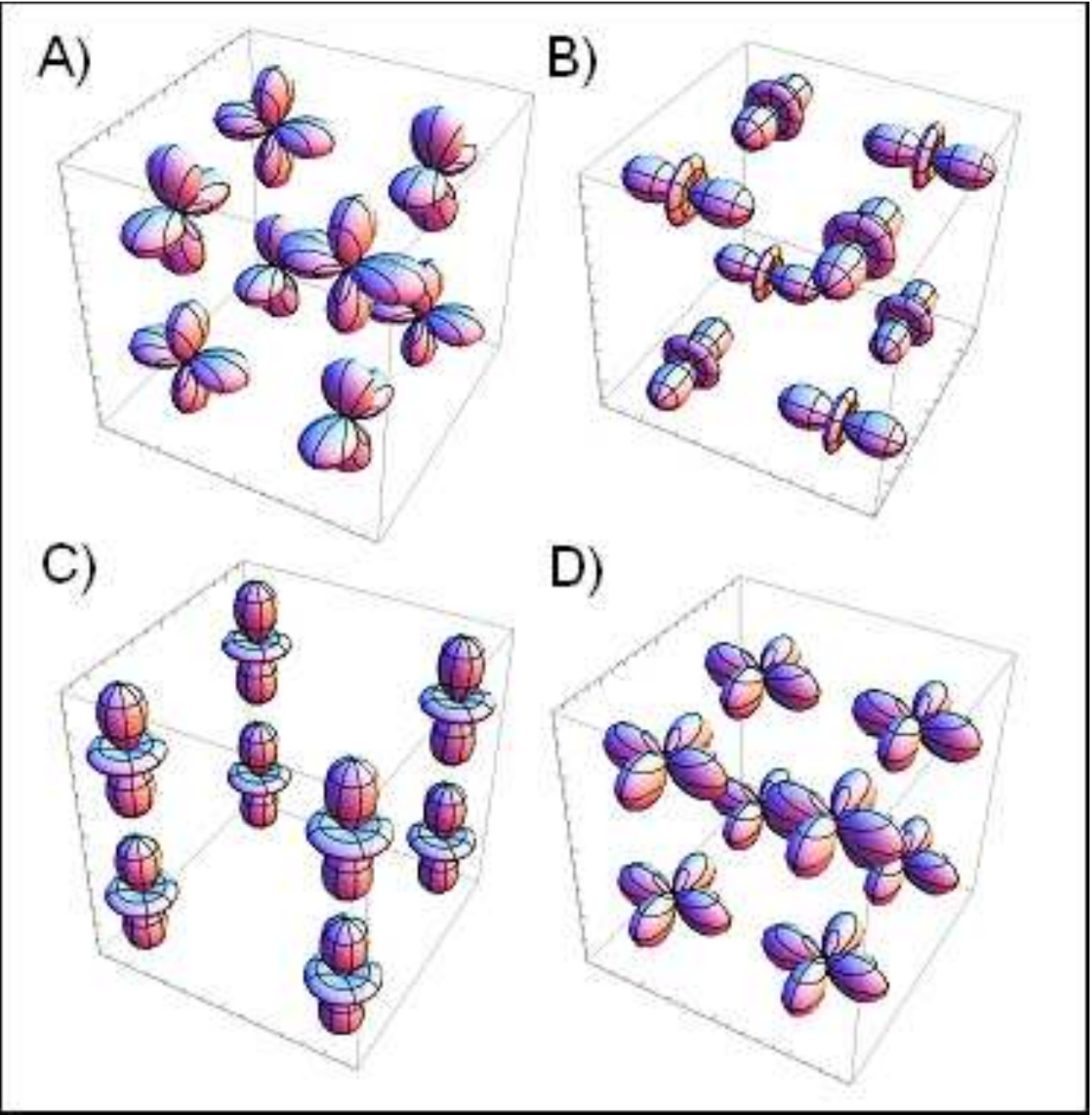}
\caption{(Color online)
Schematic view of four representative orbital configurations on
a cubic cluster:
(a) AO order with $\langle\tau^{a(b)}_i\rangle=1/2$
changing from site to site and $\langle\tau^c_i\rangle=-1/4$,
obtained for $E_z<0$,
(b) AO order with $\langle\tau^{a(b)}_i\rangle=-1/2$
changing from site to site and $\langle\tau^c_i\rangle=-1/4$,
obtained for $E_z>0$,
(c) FO order with occupied $z$ orbitals and
$\langle\tau^c_i\rangle=-1/2$ (cigar-shaped orbitals), and
(d) FO order with occupied $x$ orbitals and
$\langle\tau^c_i\rangle=1/2$ (clover-shaped orbitals).}
\label{orbs}
\end{figure}

The bilayer spin-orbital $d^9$ model Eq. (\ref{hamik}) poses a
difficult many-body problem which cannot be solved exactly. The only
simple limits are either $|E_z|\to\infty$ or $\eta\to(1/3)^-$ which we
discuss below. In the first case the dominant term is the crystal field
$\propto E_z$ and, depending on its sign, we get uniform orbital
configuration $\tau^c_i\equiv \pm 1/2$ and $\tau^{a,b}_i\equiv\mp 1/4$.
After inserting these classical expectation values into the Hamiltonian
Eq. (\ref{hamik}) we are left with the spin part which has purely
Heisenberg form.

\begin{figure}[t!]
    \includegraphics[width=8.2 cm]{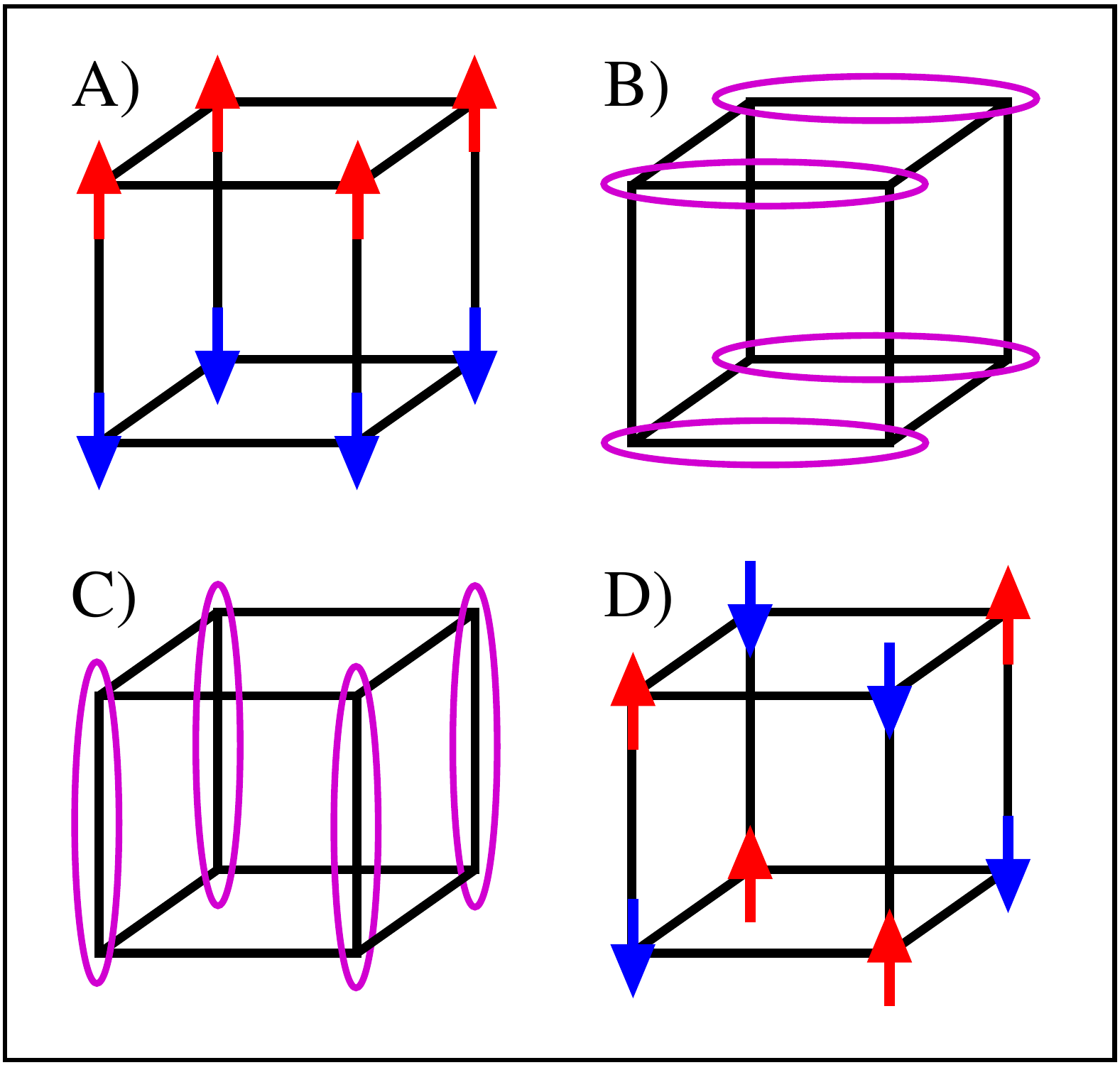}
    \caption{(Color online) Schematic view of four different spin
ordered phases on a cube realized in the $d^9$ spin-orbital
model Eq. (\ref{hamik}):
(a) $A$-AF configuration,
(b) plaquette valence bond (PVB) singlet configuration,
(c) VB$z$ phase with singlets along the $c$ axis, and
(d) $G$-AF configuration.
Arrows stand for up or down spins, oval (violet) frames indicate
singlets. Spin disordered phases with singlets on certain bonds
(b) and (c) are stabilized by particular orbital order, see Sec. III. }
    \label{4cubes}
\end{figure}

We will show below that in the bilayer geometry of the lattice the
single-site MF approximation predicts long-range ordered $G$-AF phases
at $\eta=0$ known from the 3D spin-orbital $d^9$ model,\cite{Fei97}
see Fig. \ref{4cubes}(d). For negative $E_z\to-\infty$ and FO order of
$z$ orbitals shown in Fig. \ref{orbs}(c), we get an AF coupling in the
$c$ direction and a weaker AF coupling in the $ab$ planar directions
(in the regime of small $\eta$).
For positive $E_z\to\infty$ one finds instead
the FO order of $x$ orbitals shown in Fig. \ref{orbs}(d), and two $ab$
planes decouple, so we are left with the AF Heisenberg model on two
independent 2D square lattices. In this case and the spins exhibit
either $G$-AF, see Fig. \ref{4cubes}(d) or $C$-AF order (not shown).
Ferromagnetism is obtained in the present model for any $E_z$ if $\eta$
is sufficiently large, i.e., when the superexchange is dominated by terms
proportional to $r_1$ which favor formation of spin triplets on the bonds
accompanied by AO order depicted in Figs. \ref{orbs}(a) and \ref{orbs}(b).

In what follows we will show the simplest, single-site MF
approximation of the Hamiltonian Eq. (\ref{hamik}) and the
resulting phase diagram. The Hamiltonian, originally expressed
in terms of bond operators, can be then written
in a "single-site" form given below:
\begin{eqnarray}
{\cal H}_{\rm MF}\!\!&=&\!\frac{1}{2}J\sum_{i,\gamma}\!\left\{
\tau^{\gamma}_i\tau^{\gamma}_{i+\gamma}(\chi^{\gamma}-\xi^{\gamma})
+\frac{1}{2}\tau^{\gamma}_i\xi^{\gamma}
-\frac{1}{4}(\chi^{\gamma}+\xi^{\gamma})\right\}
\nonumber\\
&-&E_z\sum_i\tau_i^c,
\end{eqnarray}
with sum running over all sites and cubic axes $\gamma=a,b,c$. Here
we adopted a shorthand notation with $i+\gamma$ meaning the nearest
neighbor of site $i$ in the direction $\gamma$.

The quantities
\begin{eqnarray}
\chi^{\gamma}=\left\{
\begin{array}{ccc}
r_1\Pi_t^{\gamma}+r_2\Pi_s^{\gamma} & if & \gamma =a,b\\
\frac{1}{2}(r_1\Pi_t^{\gamma}+r_2\Pi_s^{\gamma}) & if & \gamma =c
\end{array}
\right\},
\end{eqnarray}
and
\begin{equation}
\xi^{\gamma}=\left\{
\begin{array}{ccc}
(r_2+r_4)\Pi_s^{\gamma}& if & \gamma =a,b\\
\frac{1}{2}(r_2+r_4)\Pi_s^{\gamma}& if & \gamma =c
\end{array}
\right\},
\end{equation}
are parameters obtained by averaging over spin operators.
The coefficients $1/2$ in the $\chi^{\gamma}$ and $\xi^{\gamma}$ terms
along the $c$ axis follow from the bilayer geometry of the lattice.
We assumed that the spin order, determining $\chi^{\gamma}$ and
$\xi^{\gamma}$, depends only on the direction $\gamma$ and not on site
$i$. This is sufficient to investigate the phases with either AF or FM
long-range order. More precisely, these are spin-singlet and
spin-triplet projectors
$\Pi_{s(t)}^{\gamma}\equiv\Pi^{s(t)}_{i,i+\gamma}$ Eqs. (\ref{proje})
that are independent of $i$.
As far as only a single site is concerned the spins cannot
fluctuate at zero temperature and the projectors can
be replaced by their average values:
\begin{equation}
\label{project}
\Pi_s^{\gamma}=\frac{1}{4}-\langle{\bf S}_i\cdot{\bf S}_{i+\gamma}\rangle,
\hskip .5cm
\Pi_t^{\gamma}=\frac{3}{4}+\langle{\bf S}_i\cdot{\bf S}_{i+\gamma}\rangle.
\end{equation}
The values of the projectors depend on the assumed spin
order. Here we consider four different spin configurations:
(i) $G$-AF - antiferromagnet in all three directions shown in
Fig. \ref{4cubes}(d),
(ii) $C$-AF - antiferromagnet in the $ab$ planes with FM
correlations in the $c$ direction (not shown),
(iii) $A$-AF - AF phase with FM order in the $ab$ planes and AF
correlations in the $c$ direction depicted in Fig. \ref{4cubes}(a),
(iv) FM phase (not shown).
The numerical values of the spin projection operators in these
phases are listed in Table \ref{sord}.

\begin{table}[b!]
\caption{Mean values of triplet $\Pi_t^{\gamma}$ and singlet
$\Pi_s^{\gamma}$ projection operators Eqs. (\ref{project}) for a bond
$\langle ij\rangle$ along the axis $\gamma$ in different phases with
long-range magnetic order which occur in the MF phase diagram, see
Fig. \ref{ssmfa}.}
\begin{ruledtabular}
\begin{tabular}{ccccc}
phase & $\Pi_t^{a(b)}$& $\Pi_t^{c}$ &
        $\Pi_s^{a(b)}$& $\Pi_s^{c}$ \\
\colrule
$G$-AF    &1/2&1/2&1/2&1/2 \cr
$C$-AF    &1/2& 1 &1/2& 0  \cr
$A$-AF    & 1 &1/2& 0 &1/2 \cr
FM        & 1 & 1 & 0 & 0  \cr
\end{tabular}
\end{ruledtabular}
\label{sord}
\end{table}

After fixing spins, the MF approximation
involves the well-know decoupling
for the orbital operators:
\begin{equation}
\tau_i^{\gamma}\tau_{i+\gamma}^{\gamma}
\simeq \langle \tau_i^{\gamma} \rangle \tau_{i+\gamma}^{\gamma}+
\tau_i^{\gamma}\langle \tau_{i+\gamma}^{\gamma} \rangle -
\langle \tau_i^{\gamma} \rangle\langle \tau_{i+\gamma}^{\gamma} \rangle.
\end{equation}
The last step is to define sublattices for
the orbitals. The most reasonable choice
would be to assume AO order meaning that neighboring orbitals
are always rotated by $\pi/2$ in the $ab$ plane
with respect to each other. To implement this structure into the
MF Hamiltonian we define new direction $\bar{\gamma}$ as follows:
$\bar{\gamma}=b,a$ for $\gamma=a,b$
and $\bar{\gamma}=c$ for $\gamma=c$.
Using $\bar{\gamma}$ we can now
easily define staggered order parameters:
\begin{equation}
t_i^{\gamma}\equiv \langle \tau_i^{\gamma} \rangle =
\langle \tau_{i\pm\gamma}^{\bar{\gamma}} \rangle.
\end{equation}
The final single-site MF Hamiltonian can
be written in the same form for any site so
further on we will not use site index $i$
anymore. The desired formula is:
\begin{equation}
{\cal H}^{(0)}_{{\rm MF}}=\sum_{\gamma}\Theta^{\gamma}\tau^{\gamma}+f(t_a,t_c)
=\alpha \sigma^z+\beta\sigma^x +f(t_a,t_c),
\label{1stham}
\end{equation}
with
\begin{equation}
\Theta^{\gamma}=\frac{1}{2}\xi^{\gamma}+t^{\bar{\gamma}}
(\chi^{\gamma}-\xi^{\gamma})-E_z\delta_{\gamma c},
\end{equation}
and
\begin{equation}
f(t_a,t_c)=-\frac{1}{8}\sum_{\gamma}\left\{(\chi^{\gamma}+\xi^{\gamma})+
4t^{\gamma}t^{\bar{\gamma}}(\chi^{\gamma}-\xi^{\gamma})
\right\}.
\end{equation}
For convenience we set $J=1$; note that the energy scale can easily be
recovered by replacing $E_z$ by $E_z/J$.
As we can see the MF Hamiltonian is very simple and can be written in
terms of two Pauli matrices $\{\sigma^x,\sigma^z\}$ with
\begin{equation}
\alpha=\frac{1}{2}\left(\Theta^c-\frac{1}{2}\Theta^a-\frac{1}{2}\Theta^b\right),
\hskip .5cm
\beta=\frac{\sqrt{3}}{4}\left(\Theta^{a}-\Theta^{b}\right).
\label{abdef}
\end{equation}
Solving the $2\times 2$ eigen-problem we obtain self-consistency equations
for the order parameters $t^a$ and $t^c$:
\begin{eqnarray}
 \label{SCeq}
 t^a&=&\frac{1}{4\Delta}\left(\alpha -\sqrt{3}\beta\right),  \\
 \label{SCeq1}
 t^c&=&-\frac{1}{2\Delta}\alpha,
\end{eqnarray}
where
$\Delta=\sqrt{\alpha^2+\beta^2}$
and the ground state energy given by:
\begin{equation}
E_0=-\Delta- f(t^a,t^c).
\label{Emin}
\end{equation}

\begin{figure}[t!]
 \includegraphics[width=8.2 cm]{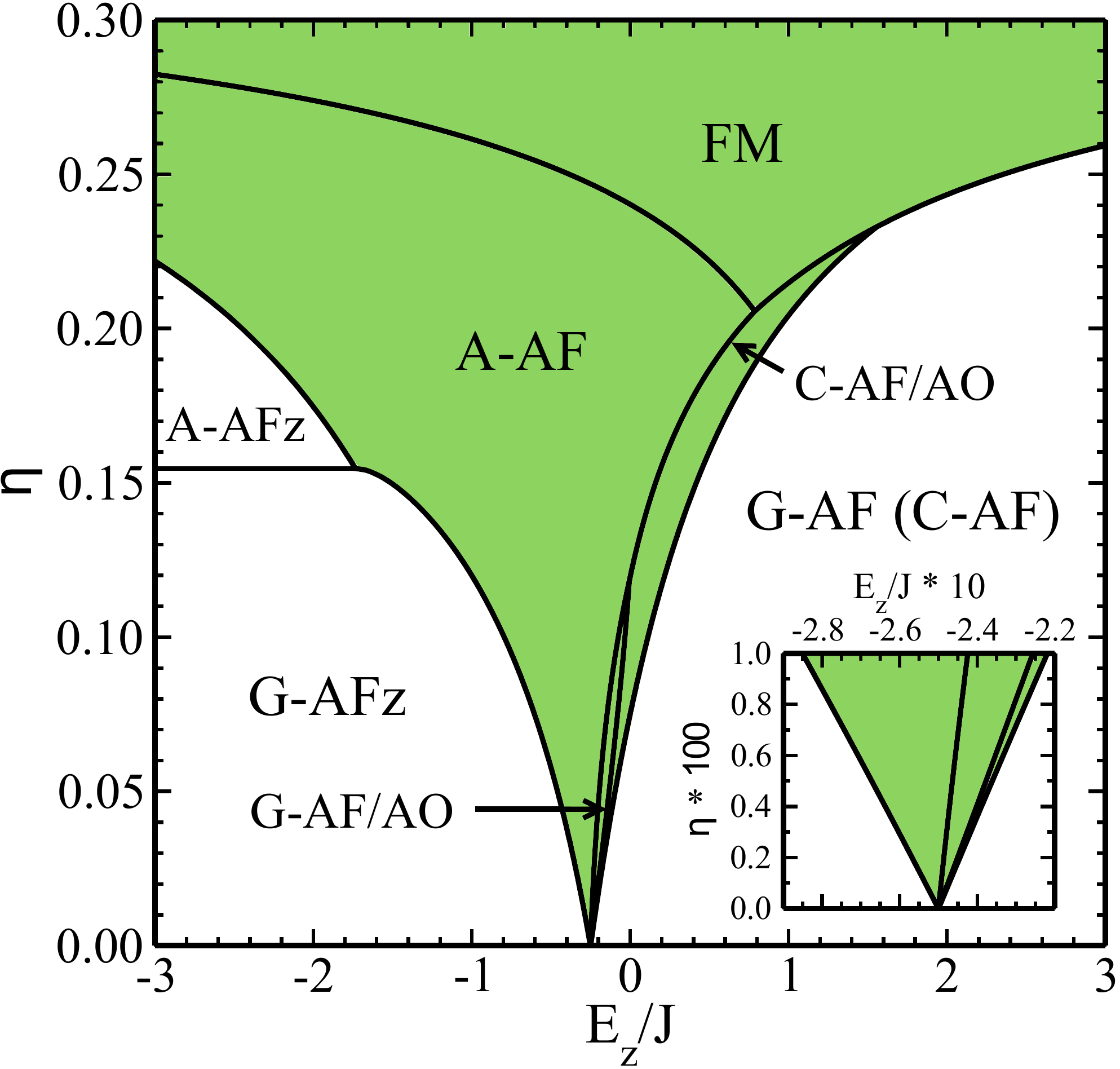}
\caption{(Color online) Phase diagram of the bilayer $d^9$ spin-orbital
model Eq. (\ref{1stham}) obtained in the single-site MF approximation
with spin and orbital MFs. In this approach the $G$-AF and $C$-AF phases
(for $E_z>-0.25$ and moderate $\eta$) have exactly the same energy.
Shaded gray (green) area indicates phases with AO order, while the
remaining states with long-range $G$-AF spin order are accompanied by
FO order. The enlarged area around the multicritical point at
$E_z=-0.25J$ and $\eta=0$ is shown in the inset.}
\label{ssmfa}
\end{figure}

The solution of self-consistency equations is
very elegant and entertaining so we are not going to present it
here and recommend it to the reader as an exercise (the results
can be next compared with those given in the Appendix).
It turns out that all four phases considered here
can appear as orbitally uniform, i.e., having FO order with orbitals
being either perfect clovers or perfect cigars everywhere,
or as phases with AO order between two sublattices.
The phase diagram presented in Fig. \ref{ssmfa} was obtained
by purely energetic consideration and shows the boarder lines
between phases with the lowest energies for
given $\eta$ and $E_z$. This diagram is
surprisingly complex taking into account the simplicity of
the single-site approach; it reveals seven different phases.
For $\eta=0$ we have only two AF phases:
(i) $G$-AF$z$ for $E_z<-1/4J$ and
(ii) $G$-AF for $E_z>-1/4J$, with a different but uniform orbital
configuration (FO order) which involves either cigar-shaped
$z$ orbitals in the $G$-AF$z$ phase, see Fig. \ref{orbs}(c), or
clover-shaped $x$ orbitals in the $G$-AF, see Fig. \ref{orbs}(d).
Because of the planar orbital configuration in the latter $G$-AF phase
one finds no interplane exchange coupling and thus this phase is
degenerate with the $C$-AF one.

For higher $\eta$ the number of phases increases abruptly by three
phases with AO configurations, as shown in the inset of Fig.
\ref{ssmfa}: the $A$-AF, $G$-AF/AO and $C$-AF/AO phase. Surprisingly,
the AO version of the $G$-AF phase is connected neither to
$z$ nor to $x$ FO order in an
antiferromagnet, excluding the multicritical point
at $(E_z/J,\eta)=(-0.25,0)$, and disappears completely for
$\eta\approx 0.118$. The $C$-AF/AO phase stays
on top of uniform $G$($C$)-AF phase, lifting the degeneracy
of the above phases at relatively large $\eta$ and then gets
replaced by the FM phase which always coexists with AO order.
One can therefore conclude that the $G$($C$)-AF degeneracy is most
easily lifted by turning on the orbital alternation.

On the opposite side of the diagram the $G$-AF$z$ phase is completely
surrounded by $A$-AF phases: for $\eta>(2/\sqrt{3}-1)$ the $G$-AF$z$
phase turns into orbitally uniform $A$-AF$z$
independently of the value of $E_z$ (interorbital triplet excitations
dominate then on the bonds in the $ab$ planes), and for smaller $\eta$
into the $A$-AF phase with AO order. In the $A$-AF phase the AF
correlations in the $c$ direction survive despite the overall FM
tendency when $\eta$ grows. This follows from the orbitals'
elongation in the $c$ direction present for $E_z<0$, which would cause
interplane singlets formation if we were not working in single-site MF
approximation, see Sec. III. In the present case it favors either the
$G$-AF$z$ or $A$-AF($z$) configuration with uniform or alternating
orbitals depending on the values of $E_z/J$ and $\eta$. Finally, the
FM phase is favorable for any $E_z$ if only $\eta$ is sufficiently
close to $1/3$ which only confirms that the single-site MF
approximation is sound and not totally wrong with this respect.

The central part of the presented diagram is the most frustrated one
judging by the number of competing phases with long-range spin order.
This behavior is consistent with that found in the 3D spin-orbital
$d^9$ model in the regime of $E_z\simeq 0$ and finite $\eta$.\cite{Fei97}
Four of these phases could be expected by looking at the phase diagram
of the 3D model: two $G$-AF phases,
the $A$-AF phase and the FM phase.\cite{Fei97}
Note, however, that in the phases stable in the central part of the
phase diagram, namely in the $A$-AF, $A$-AF/AO and FM phase, the
occupied orbitals alternate. While the FM phase is not surprising in
this respect and obeys the Goodenough-Kanamori rule of having FM spin
order accompanied by the AO order, in the $A$-AF one finds an example
that both spin and orbital order could in principle alternate between
the two $ab$ planes. This finding suggests that in this central part of
the phase diagram one may expect either other VB-type phases
or even states with more complex spin-orbital disorder. Such
ordered or disordered phases require a more sophisticated approach,
either variational wave functions,\cite{Fei97,Nor08} or the embedded
cluster approach which we explain below in Sec. \ref{sec:mfa}

\subsection{Cluster mean--field Hamiltonian}
\label{sec:mfa}

\begin{figure}[t!]
    \includegraphics[width=8.2cm]{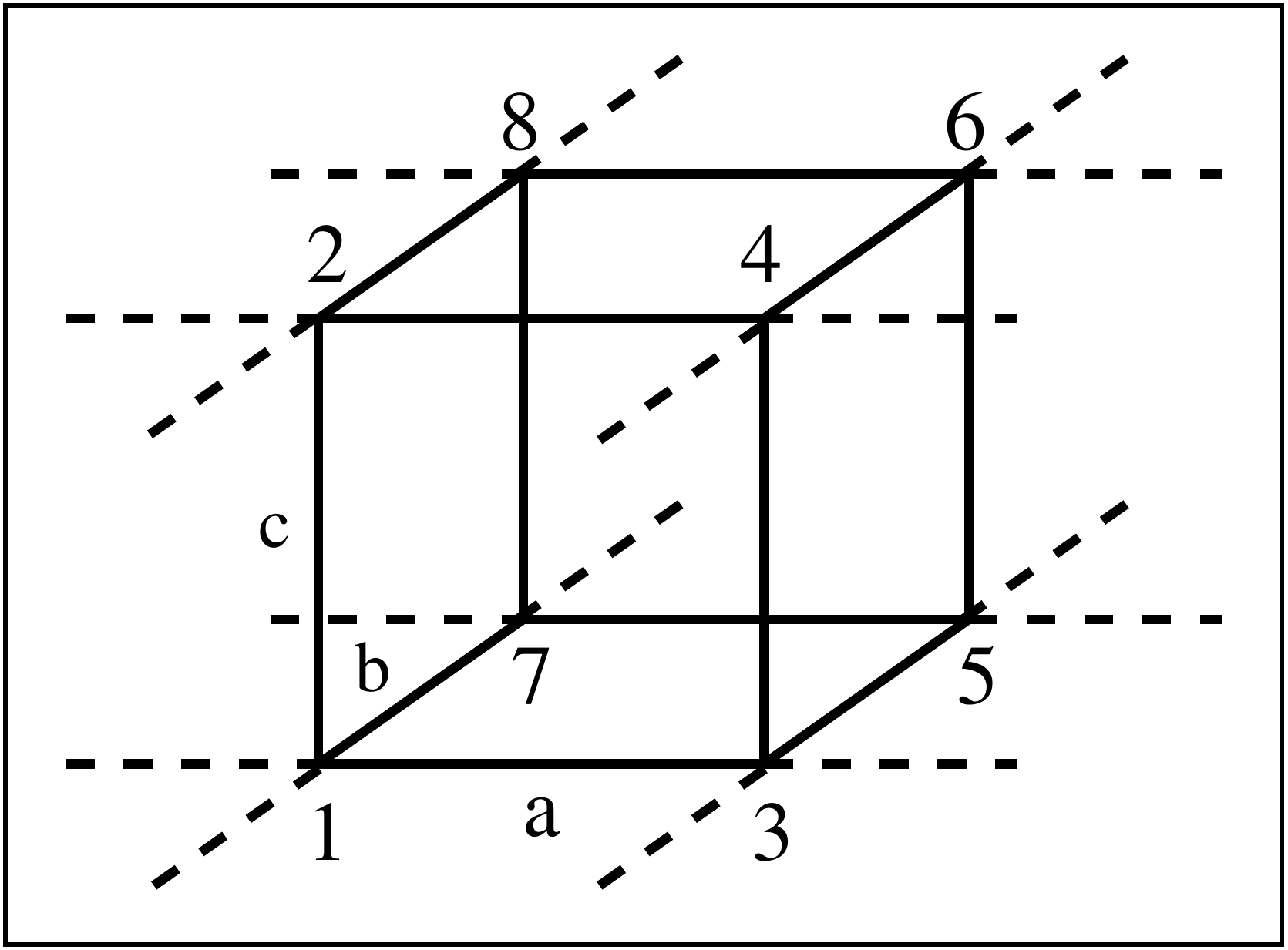}
\caption{Schematic view of the cluster used in the Bethe--Peierls--Weiss
MF approach of Sec. \ref{sec:mfa}. Vertices $i=1,\cdots,8$
             and directions $\gamma=a,b,c$ are marked in the figure,
             and dashed lines stand for the outgoing MF
             interactions in the $a$ and $b$ direction.}
    \label{cub}
\end{figure}

Now we introduce a more sophisticated approach which goes beyond the
single-site MF approximation of Sec. \ref{sec:ssa}. In what follows we
use a cluster MF approach with a cube depicted in Fig. \ref{cub}.
It contains eight sites coupled to its neighbors along the bonds in
$ab$ planes by the MF terms. This choice is motivated by the form
of the Hamiltonian with different interactions along the bonds in three
different directions
--- the cube is the smallest cluster which does not break the symmetry
between the $a$ and $b$ axes and contains equal numbers of $a$, $b$ and
$c$ bonds. After dividing the entire bilayer square lattice into
identical cubes which cover the lattice, the Hamiltonian (\ref{hamik})
can be written in a cluster MF form as follows,
\begin{eqnarray}
{\cal H}_{\rm MF}=\sum_{m\in{\cal C}}({\cal H}^{\rm int}_m+{\cal
H}^{\rm ext}_m)\,, \label{mfham}
\end{eqnarray}
where the sum runs over the set of cubes ${\cal C}$, with
individual cube labeled by $C_m\in{\cal C}$. Here ${\cal H}^{\rm int}_m$
contains all bonds from ${\cal H}_m$ belonging to the cube $C_m$ and
the crystal field terms $\propto E_z$, i.e., it depends only on
the operators on the sites inside the cube, while ${\cal H}^{\rm ext}_m$
contains all bonds outgoing from the cube $m$ and connecting
neighboring clusters, making them correlated.

The basic idea of the cluster MF approach is to approximate
${\cal H}^{\rm ext}_m$ by $\tilde{{\cal H}}^{\rm ext}_m$ containing
only operators from the cube $m$. This can be accomplished in many
different ways depending on which phase we wish to investigate. Our
choice will be to take $\tilde{{\cal H}}^{\rm ext}_m$ of the following
form:
\begin{equation}
\tilde{{\cal H}}^{\rm ext}_m=\frac{1}{2}\!\sum_{\gamma=a,b\atop i\in C_m}
\left\{S^z_ia^{\gamma}_i+S^z_i\tau^{\gamma}_ib^{\gamma}_i
+\tau^{\gamma}_ic^{\gamma}_i+d^{\gamma}_i \right\},
\label{htild}
\end{equation}
containing spin field $S^z_i$ breaking SU(2) symmetry, orbital field
$\tau^{\gamma}_i$ and spin--orbital field $S^z_i\tau^{\gamma}_i$.
Coefficients $\{a^{\gamma}_j,b^{\gamma}_j,
c^{\gamma}_j,d^{\gamma}_j\}$ are the Weiss fields and should be
fixed self--consistently depending on $E_z$ and $\eta$. Our
motivation for such expression is simple: if orbital degrees of
freedom are fixed then the problem reduces to the Heisenberg model
which has long--range ordered AF phase --- that is why we take
$S^z_i$ field, the orbitals are present in the Hamiltonian so
taking $\tau^{\gamma}_i$ is the simplest way of treating them on
equal footing to describe possible orbital order. Finally, we
introduce also spin-orbital field $S^z_j\tau^{\gamma}_j$ because we
believe that in some phases spins and orbitals alone do not suffice
to describe the symmetry breaking and these operators can act
together.

The standard way to go on is to write self--consistency equations
for the Weiss fields. This can be done in a straightforward
fashion: we take the operator products from ${\cal H}^{\rm ext}_m$
and divide them into a part depending only on $O_i$ operators from
the cube $m$, and on $O_j$ ones --- from a neighboring cube $n$.
Then we use the well known MF decoupling for such operator
products,
\begin{eqnarray}
O_iO_j
&\approx &\langle O_i \rangle O_j+O_i\langle O_j \rangle-
\langle O_i \rangle\langle O_j \rangle  \nonumber\\
&=&O_i\langle O_j \rangle-\frac{1}{2}\langle O_i \rangle\langle
O_j \rangle+ O_j\langle O_i \rangle-\frac{1}{2}\langle O_i
\rangle\langle O_j \rangle\,, \nonumber\\ \label{deco}
\end{eqnarray}
and write it in a symmetric way. Now the first two terms can be
included into $\tilde{{\cal H}}^{\rm ext}_m$ and the last two into
$\tilde{{\cal H}}^{\rm ext}_n$. This procedure can be applied to all
operator products in ${\cal H}^{\rm ext}_m$ and full
$\tilde{{\cal H}}^{\rm ext}_m$ can be recovered in the form
given by Eq. (\ref{htild}). Repeating this for all clusters
leads to a Hamiltonian describing a set of commuting cubes
interacting in a self--consistent way. After using Eq.
(\ref{deco}) on the Hamiltonian Eq. (\ref{hamik}) we obtain the
formulas for the Weiss fields:
\begin{eqnarray}
a^{\gamma}_i&=&\frac{1}{2}(r_2+r_4)u^{\gamma}_i+\frac{1}{4}(r_2-r_1)s^{\gamma}_i,
  \\
b^{\gamma}_i&=&-(r_4+r_1)u^{\gamma}_i-\frac{1}{2}(r_2-r_1)s^{\gamma}_i,
  \\
c^{\gamma}_i&=&\frac{1}{4}(3r_1-r_4)t^{\gamma}_i+\frac{1}{8}(r_2+r_4),
  \\
d^{\gamma}_i&=&-\frac{1}{2}(r_1+r_4)u^{\gamma}_iu^{\gamma}_{m,i}
-\frac{1}{4}(r_2-r_1)(s_iu^{\gamma}_i+s^{\gamma}_iu^{\gamma}_{m,i})
\nonumber \\
&-&\frac{1}{16}(r_2+r_4)(t^{\gamma}_{m,i}-t^{\gamma}_i)
+\frac{1}{8}(r_4-3r_1)t^{\gamma}_it^{\gamma}_{m,i} \nonumber \\
&-&\frac{1}{32}(3r_1+2r_2+r_4),
\end{eqnarray}
where the order parameters at site $i$ are:
\begin{eqnarray}
s_i&\equiv &\left\langle S^z_i\right\rangle,\\
t^{\gamma}_{m,i}&\equiv &\left\langle\tau^{\gamma}_i\right\rangle,\\
u^{\gamma}_{m,i}&\equiv &\left\langle
S^z_i\left(\frac{1}{2}-\tau^{\gamma}_i\right)\right\rangle.
\label{mdef}
\end{eqnarray}
Note that $\{s_i,t^{\gamma}_{m,i}, u^{\gamma}_{m,i}\}$
are the mean values of operators at site $i$
belonging to the cluster $m$, and
$\{s^{\gamma}_i,t^{\gamma}_{i}, u^{\gamma}_{i}\}$ are the mean
values of the same operators at sites neighboring with $i$ in the
direction $\gamma$. The geometry of a bilayer implies that each
site $i$ has one neighbor along the axis $a$ and another one along
the axis $b$, and these sites belong to different cubes.

The next crucial step is to impose a condition that
$\{s^{\gamma}_i,t^{\gamma}_{i}, u^{\gamma}_{i}\}$ are related to
the order parameters obtained on the internal sites of the considered
cluster. The simplest solution is to assume that all clusters have
identical orbital configuration; $t^{\gamma}_{i}=t^{\gamma}_{m,i}$,
spin configuration is in agreement with a type of global magnetic order
we want to impose; $s^{\gamma}_{i}=\pm s_{i}$ and spin orbital
configuration is as if spin and orbitals were factorized, i.e.,
$u^{\gamma}_{i}=\pm u^{\gamma}_{m,i}$.
This solution has one disadvantage: if $a$ or $b$ direction is
favored in the orbital configuration of the cube then this broken
symmetry will propagate through whole lattice which is
contradictory with the form of the Hamiltonian Eq. (\ref{hamik}).
That is why it is better to assume that two neighboring cubes can
differ in orbital (and spin-orbital) configuration by the
interchange of $a$ and $b$ direction, i.e.,
\begin{eqnarray} \label{szf}
s^{\gamma}_i=\pm s_i,\hskip .7cm \label{tzf}
t^{\gamma}_{i}=t^{\bar{\gamma}}_{m,i},\hskip .7cm \label{stf}
u^{\gamma}_{i}=\pm u^{\bar{\gamma}}_{m,i} \label{legs}
\end{eqnarray}
with $\bar{\gamma}$ being the complementary direction in the $ab$ plane
to $\gamma$, i.e., $(\gamma,\bar{\gamma})=(a,b),(b,a)$. This relation
gives the same results as the previous one in case when the $(a,b)$
symmetry in the cube is not broken, but keeps the whole system $(a,b)$
symmetric in the other case. Here we again treat the spin-orbital field
as factorized but surprisingly it turns out that this does not prevent
spin-orbital entanglement to occur, see below. We have also tried to
impose relations between $u^{\gamma}_{i}$ and $u^{\gamma}_{m,i}$ which
have nothing to do with spin and orbital sectors alone but this only
resulted in the lack of convergence of self--consistency iterations.

\subsection{Self--consistent iterative procedure}
\label{sec:ite}

The self--consistency equations cannot be solved exactly because
the effective cluster Hilbert space is too large even if we use total
${\cal S}^z$ conservation in the considered cluster $m$ (then the
largest subspace dimension is $d=17920$) and because of their
non--linearity. The way out is to use Bethe--Peierls--Weiss method,
i.e., set certain initial values for the order parameters
$\{s_i,t^{\gamma}_{m,i},u^{\gamma}_{m,i} \}$ and next employ Lanczos
algorithm to diagonalize ${\cal H}_{\rm MF}$ Eq. (\ref{mfham}). Below
we present results obtained by self-consistent calculations of phases
with broken symmetry or with spin disorder. In order to determine
the ground state one recalculates mean values of spin, orbital and
spin-orbital fields given by Eqs. (\ref{stf}) and determines new
order parameters. This procedure is continued until convergence
(of energy and order parameters) is reached. This process can be very
slow due to the number of order parameters which is $24$ (three per
site) for the cube but we have overcome this problem by {\it
imposing} certain symmetry breaking on the cluster. We implement it in
the following way: after each iteration we calculate
$\{s_i,t^{\gamma}_{m,i},u^{\gamma}_{m,i}\}$ only for one site
$i=1$ and the remaining coefficients are fixed assuming certain
symmetries of the phase we are searching for.

For simplicity let us enumerate the vertices $i=1,\cdots,8$ in the cubic
cluster as shown in Fig. \ref{cub}. To obtain $G$-AF phases we assume
that:
\begin{eqnarray}
s_i=\left\{
\begin{array}{c}
s_1 \;\;\;if\;\;\; i\in A\\
-s_1 \;\;if\;\; i\in B
\end{array}
\right\}
\end{eqnarray}
for a two-sublattice structure, where $A=\{1,4,5,8\}$ and
$B=\{2,3,6,7\}$. In FM case it is enough to put $s_i\equiv s_1$
and in case of FM order within the planes and AF between them (in
the $A$-AF phase) we use instead: $s_i= (-1)^{i-1}s_1$. In the
orbital sector we can impose a completely uniform configuration with
$t^{\gamma}_{m,i}\equiv t^{\gamma}_{m,1}$, which can however
lead to non--uniform configuration of the whole system because
neighboring clusters are rotated by $\pi/2$ with respect to each
other, or we can produce a phase with AO order taking:
\begin{eqnarray}
t^{\gamma}_{m,i}=\left\{
\begin{array}{c}
t^{\gamma}_{m,1} \;\;\;if\;\; i\in A\\
t^{\bar{\gamma}}_{m,1} \;\;if\;\; i\in B
\end{array}
\right\}
\end{eqnarray}
with $(\gamma,\bar{\gamma})=(a,b),(b,a)$. Other choices would be to
take the above equation either with $A=\{1,2,5,6\}$ and $B=\{3,4,7,8\}$
or with $A=\{1,3,5,7\}$ and $B=\{2,4,6,8\}$. More generally
speaking, every choice of orbital sublattices is good as long as
the total MF wave function does not violate the symmetry between
directions $a$ and $b$. The sublattices for spin-orbital field are
constructed as if $u^{\gamma}_{m,1}$ could be expressed as
$u^{\gamma}_{m,i}=s_i(\frac12-t^{\gamma}_{m,i})$.

\section{Phase diagrams}
\label{sec:phd}

\subsection{Disentangled spin and orbital operators}
\label{sec:dis}

\begin{figure}[t!]
    \includegraphics[width=8.2cm]{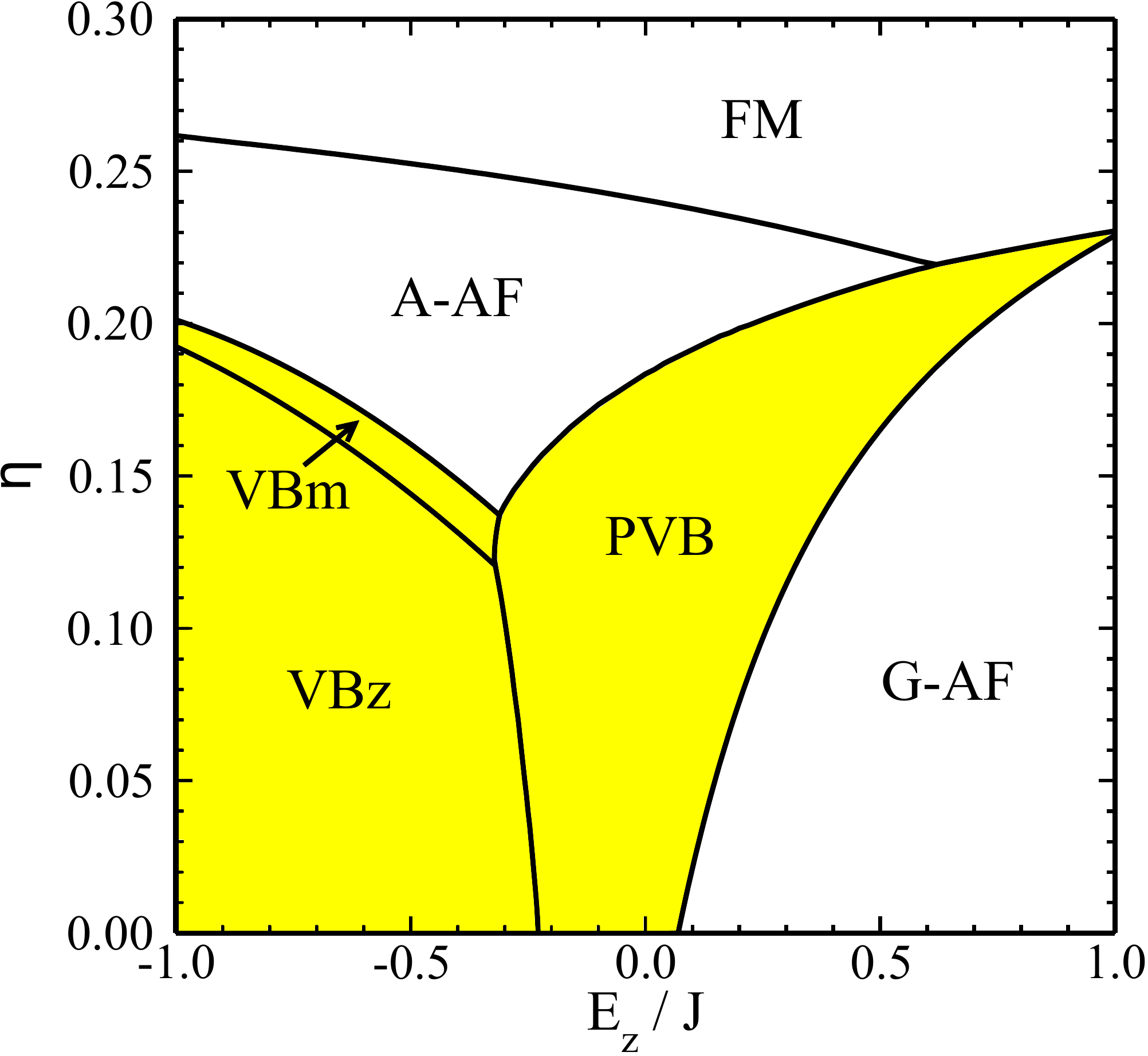}
\caption{(Color online) Phase diagram of the $d^9$ spin-orbital
superexchange model Eq. (\ref{mfham}) obtained using the cluster
method for an embedded cube with factorized spin-orbital operators.
Valence-bond phases with spin disorder are stable in the shaded area.}
    \label{diagns}
\end{figure}

The zero--temperature phase diagram of the present bilayer $d^9$
spin-orbital model Eq. (\ref{hamik}) depends on parameters
$(E_z,\eta)$, and was obtained by comparing ground state energies
for different sublattices formed by $\{s_i,t^{\gamma}_{m,i},
u^{\gamma}_{m,i}\}$ MFs. In this way we determined the ground state
with the lowest energy and its order parameters.
We begin with the phase diagram of Fig. \ref{diagns} obtained by
assuming that spin-orbital operators may be factorized into spin and
orbital parts, i.e., $u^{\gamma}_{m,i}\equiv s_i(1/2-t^{\gamma}_{m,i})$
or:
\begin{eqnarray}
\langle S^z_i \tau^{\gamma}_i\rangle \equiv
\langle S^z_i\rangle \langle \tau^{\gamma}_i\rangle.
\end{eqnarray}
Next we report the phase diagram (in Sec. \ref{sec:sof}), where we
include $u^{\gamma}_{m,i}$ calculated following the definition in Eq.
(\ref{mdef}). Comparing these two schemes allows us to determine which
phases cannot exist without spin-orbital entanglement.

The low--$\eta$ part of the diagram in Fig. \ref{diagns} is
dominated by three phases: VB$z$ for negative $E_z$, PVB for $E_z$
close to zero and $G$-AF for positive $E_z$. The VB$z$ phase with
ordered interlayer valence bonds for occupied $z$ orbitals and
spin singlets, see Fig. \ref{4cubes}(c), has replaced the $G$-AFz phase
obtained before in Sec. \ref{sec:ssa}. Both phases exhibit uniform
FO order, i.e., $t^c_{m,i}$ is close to $-1/2$ for all $i$ which means
that orbitals take the shape of cigars aligned along the $c$ bonds, see
Fig. \ref{orbs}(c). One finds that quantum fluctuations which could be
included within the present approach select the VB$z$ phase and
magnetization vanishes due to the singlets formation.
For higher values of $E_z\simeq 0$ also a different phase is found:
the plaquette VB (PVB) phase with singlets formed on the bonds in $a$
or $b$ direction of the cluster, see Fig. \ref{4cubes}(b). This phase
breaks the $a$--$b$ symmetry of the model locally but the global
symmetry is preserved thanks to the $\pi/2$ rotation of neighboring
clusters (see Eq. (\ref{legs})). The orbitals are again uniform within
the cluster with $t^a_{m,i}$ or $t^b_{m,i}$ close to $-1/2$, meaning
that they take shape of cigars pointing in the direction of the
singlets. For high positive values of $E_z$ the ground state is the
$G$-AF phase with long--range AF order and FO order of $x$ occupied
orbitals, i.e., $t^c_{m,i}$ close to $1/2$, see Fig. \ref{orbs}(d).
This means that orbitals are indeed of the $x$ type and take shape of
four--leaf clovers in the $ab$ plane with lobes pointing along $a$ and
$b$ directions which makes the two planes very weakly coupled.

The FO order in the VB$z$ and $G$-AF phases agrees with the
limiting configurations for $E_z\to\pm\infty$ described earlier.
The first of them is a quantum phase with local singlets, in
contrast to the $G$-AF$z$ one found already in Sec. \ref{sec:ssa}
and in the 3D spin-orbital $d^9$ model.\cite{Fei97} If we consider
now the VB$z$ phase and increase $\eta$, we pass through the VBm
phase (where "m" stands for mixed orbital configuration) and reach
the $A$-AF phase with non--zero global magnetization such that
spins order ferromagnetically in the $ab$ planes and
antiferromagnetically between them (along $c$ axis), see Fig.
\ref{4cubes}(a). We believe that this regime of the phase diagram
is of relevance for the spin and orbital correlations in
K$_3$Cu$_2$F$_7$ and discuss it also in Sec. \ref{sec:summa}. The
orbital order is of the AO type with $t^c_{m,i}$ close to zero,
positive or negative depending on $E_z$, see Figs. \ref{orbs}(a)
and \ref{orbs}(b). The VBm phase occurs when the orbitals in the
VB$z$ phase start to deviate from the uniform configuration and
ends when the global magnetization appears, accompanied by the
change of the orbital order. The first transition is of second
order, being the only second order phase transition in this
diagram of Fig. \ref{diagns}.

The presence of both $A$-AF phases on top of the VB$z$ can be understood
qualitatively as follows: in the VB$z$ phase AF spin coupling is strong
only within the singlets, so if $\eta$ is increased the weak in-plane
spin correlations can easily switch to FM ones, while AF correlations
will still survive between the planes. The last phase of the diagram is
the FM phase with AO order, similar to the AO order in the $A$-AF phase.
Due to the absence of thermal and quantum fluctuations the
magnetization in this phase is constant and maximal. The FM phase
appears for any $E_z$, if only $\eta$ is sufficiently close to $1/3$,
which agrees qualitatively with the previous discussion of the exact
limiting configurations and with the phase diagram found before in
the single-site MF approach, see Fig. \ref{ssmfa}.

Comparing Fig. \ref{diagns} to the MF phase diagram of Fig.
\ref{ssmfa} we can immediately recognize the main difference: the
existence of the VB$z$ and PVB phases. These phases contain spin
singlets on the bonds and do not follow from the single-site MF
approach. Another difference is the lack of sharp transitions between
AO and FO order within one phase; these transitions are smoothened by
spin fluctuations absent in the single-site MF and perfect FO
configurations are now available only for extremely high
values of $|E_z|$.

\subsection{Phase diagram with spin-orbital field}
\label{sec:sof}

\begin{figure}[t!]
    \includegraphics[width=8.2cm]{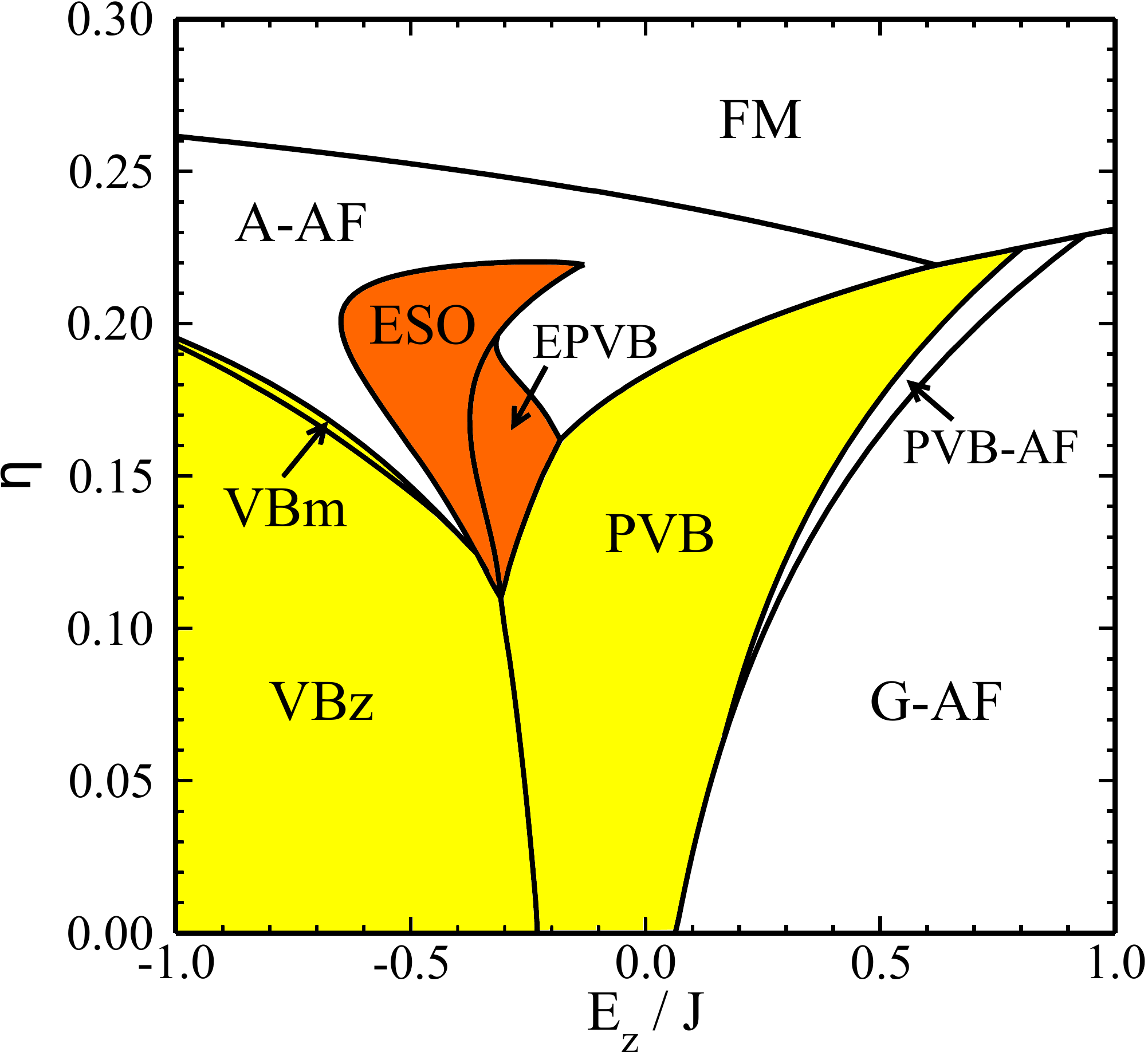}
\caption{(Color online) The phase diagram of the cluster MF Hamiltonian
Eq. (\ref{mfham}) of the $d^9$ spin-orbital model for a bilayer,
with independent spin, orbital and spin-orbital mean fields.
Valence-bond phases with spin disorder are stable in the light shaded
(yellow) area, and phases with spin-orbital entanglement are
indicated by dark gray (orange) shading.}
    \label{diag}
\end{figure}

When the spin-orbital MF is not factorized but calculated according to
its definition given in Eq. (\ref{mdef}), one finds the phase diagram
displayed in Fig. \ref{diag}.
We would like to emphasize that this non-factorizability cannot be
included within the single-site MF approach because there all spin
fluctuations are absent. Of course, one can imagine that we
take the $S^z_i\langle S^z_{i+\gamma}\rangle$ decoupling
in the pure-spin sector and $S^x_i\langle S^x_{i+\gamma}\rangle$
decoupling in the spin-orbital sector of
the Hamiltonian Eq. (\ref{hamik}) leading to the
fluctuating spins but this would break both the magnetization
conservation and homogeneity of the spin-spin
interactions included into the Kugel-Khomskii model.

In addition to the phases obtained in the phase diagram of Fig.
\ref{diagns}, we get here also the following phases: ESO, EPVB and
PVB-AF (the VBm phase is still stable between the VB$z$ and $A$-AF ones
but has much smaller area). The first two above
phases are formed in the highly frustrated region of the phase diagram
where both $E_z$ and $\eta$ are moderate. ESO stands for entangled
spin--orbital phase and is characterized by relatively high values of
spin-orbital order parameters, especially for high $\eta$ values when
other order parameters are close to zero. This phase contains singlets
along the bonds parallel to the $c$ axis, its magnetization vanishes and
the orbital configuration is uniform. One can say that this is the VB$z$
phase with weakened orbital order transformed into uniform spin-orbital
order for the same spin and orbital sublattices. EPVB stands for
entangled PVB phase and resembles it, but has in addition finite
non--uniform spin-orbital fields, and weak global AF order.
A different type of phase with spin-orbital entanglement is the
PVB-AF phase connecting PVB and $G$-AF in a smooth
(as it will be shown below) way but only if $\eta$ is large enough.
In contrast to the direct PVB-to-$G$-AF transition, passing through
the PVB-AF involves second order phase transitions and the same
happens in case of the EPVB connecting the ESO and PVB phases.
Similarly to the previous diagram, the transition from the VB$z$ to VBm
phase is of the second order while the other transitions produce
discontinuities in order parameters (see Sec. \ref{sec:ops}) and
correlation functions (see Sec. \ref{sec:corr}).

Finally, we should also point out that the $G$-AF/$C$-AF
degeneracy found in Fig. \ref{ssmfa} is lifted in the cluster
approach and the $C$-AF phase does not appear in any of the two
phase diagrams presented in Figs. \ref{diagns} and \ref{diag}.
Another interesting feature of the phase diagrams are points of
high degeneracy where different phases have the same ground-state
energies. In case of the single-site MF diagram this quantum
critical point is found at $(E_z=-1/4J,\eta=0)$, where six phases
meet. The use of cluster MF method which includes singlet phases
lifts this point upwards along the border line between VB$z$ and
PVB to $(E_z,\eta)\approx (-0.3J,0.11)$ in case of Fig.
\ref{diag}. This means that singlet formation acts against
interaction frustration caused by Hund's exchange coupling and
moves the most frustrated region of phase diagram to high-$\eta$
regime. This shows once again that the simple single-site approach
is not sufficient to describe correctly the properties of the
bilayer $d^9$ spin-orbital model.

\section{The order parameters}
\label{sec:ops}

\begin{figure}[t!]
    \includegraphics[width=8cm]{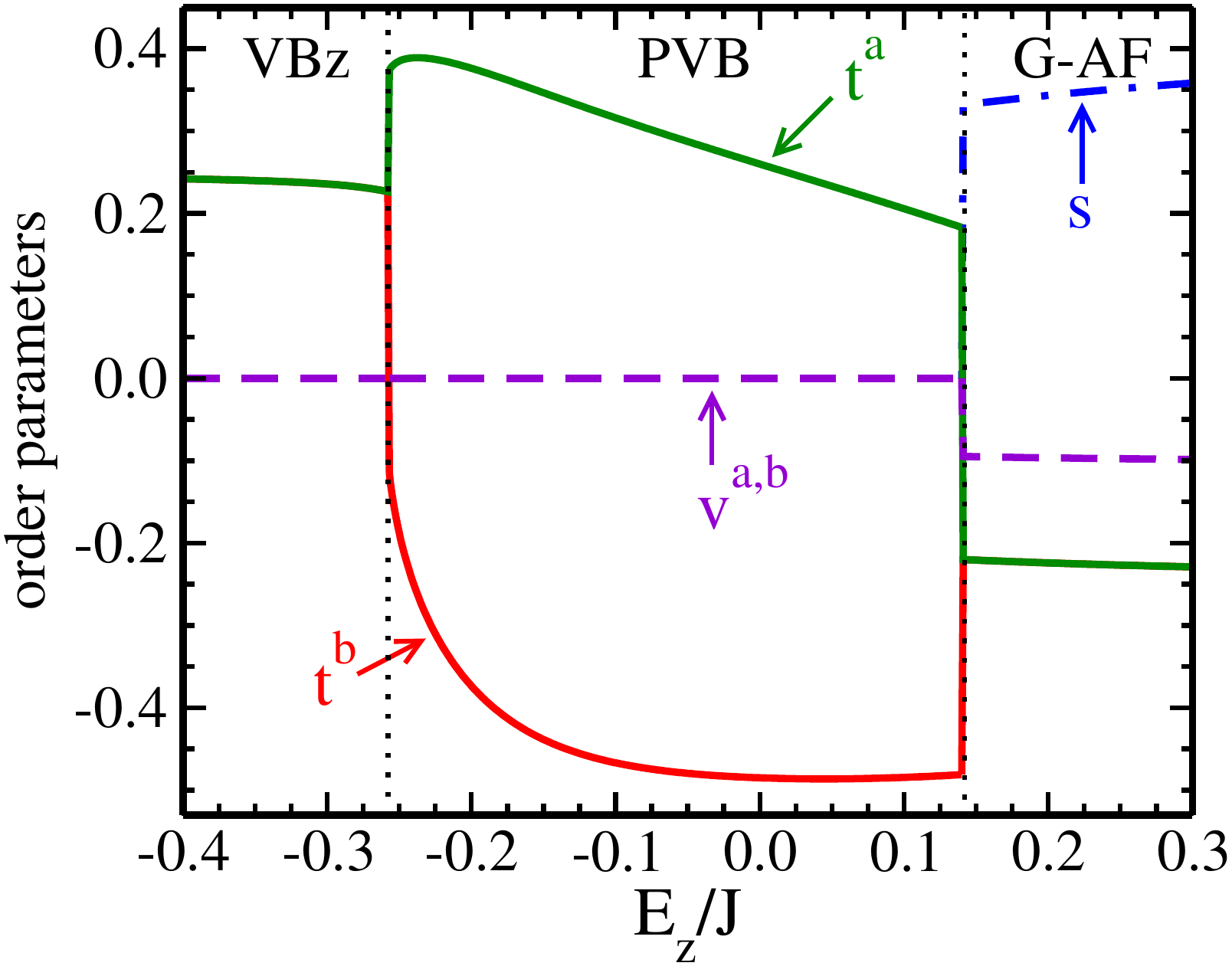}
    \caption{(Color online) Order parameters $\{s,t^{a,b},
                  v^{a,b}\}$ for $\eta =0.05$ and $-0.4<E_z/J<0.3$ in the
                  VB$z$, PVB and $G$-AF phase, from left to right.}
    \label{w1}
\end{figure}

\begin{figure}[t!]
    \includegraphics[width=8cm]{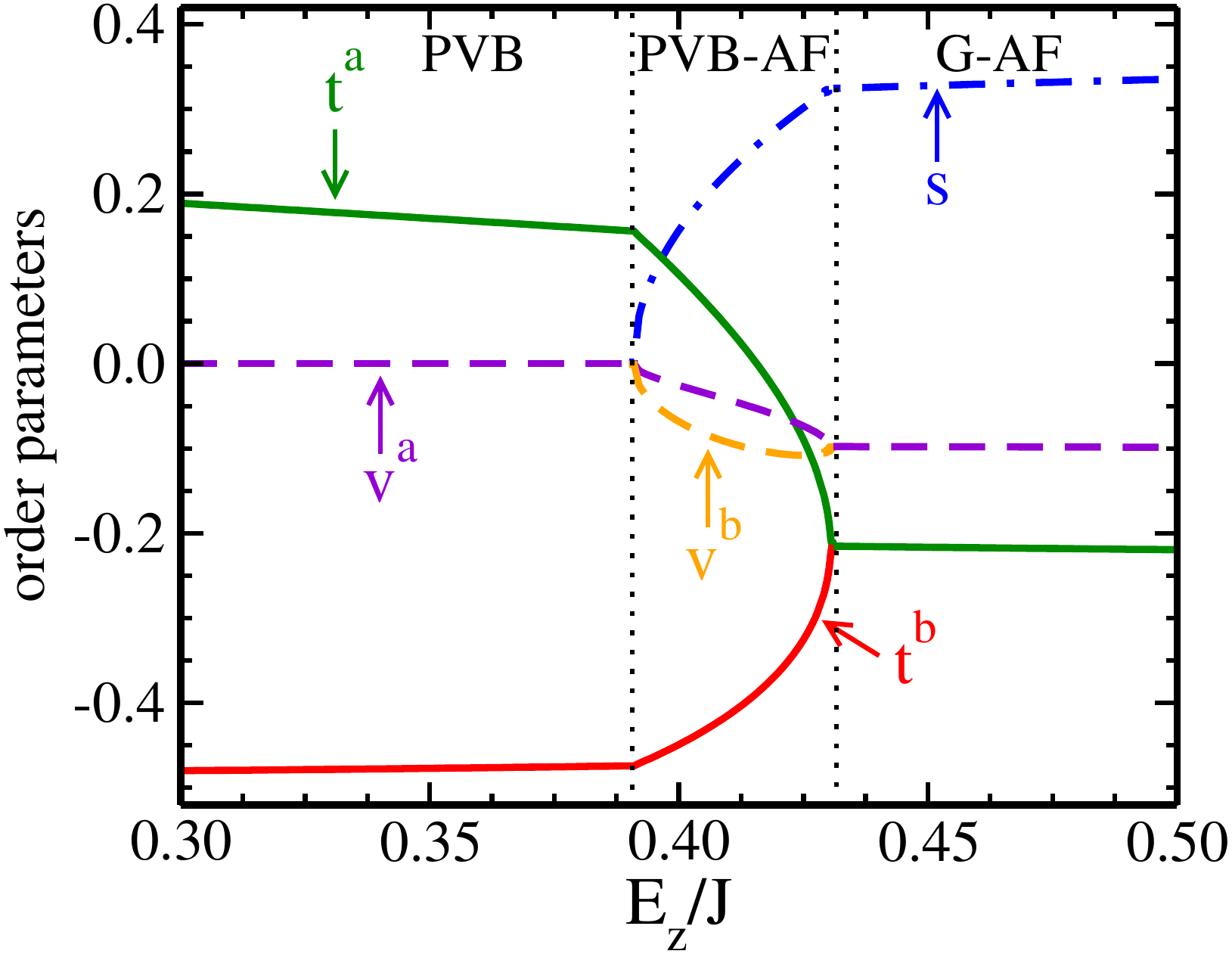}
    \caption{(Color online) Order parameters $\{s,t^{a,b},
                  v^{a,b}\}$ for $\eta =0.15$ and $0.3<E_z/J<0.5$ in the
                  PVB, PVB-AF and $G$-AF phase, from left to right.}
    \label{w2}
\end{figure}

\begin{figure}[t!]
    \includegraphics[width=8cm]{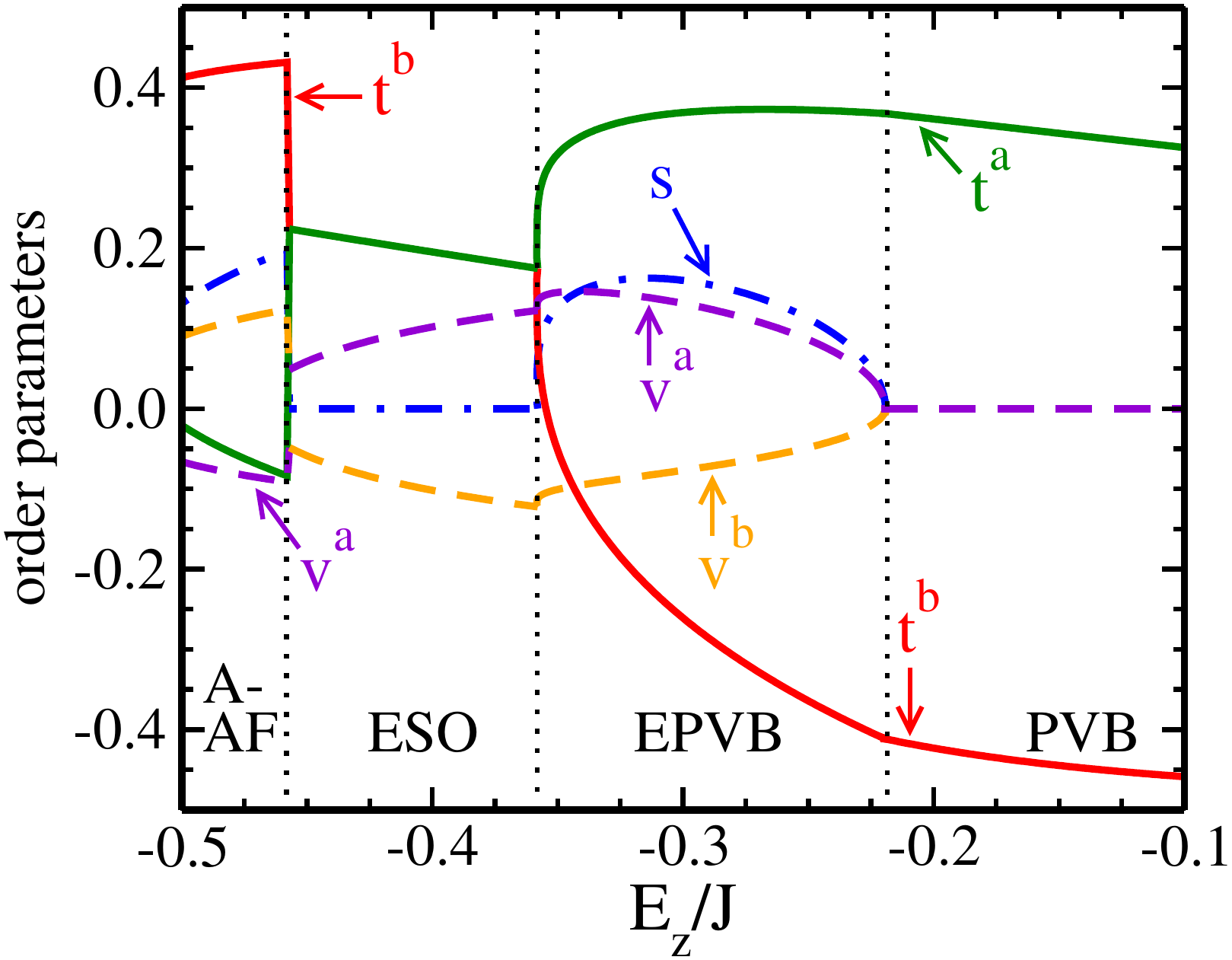}
    \caption{(Color online) Order parameters $\{s,t^{a,b},
                  v^{a,b}\}$ for $\eta =0.15$ and $-0.5<E_z/J<-0.1$ in the
                  $A$-AF, ESO, EPVB and PVB phases, from left to right.}
    \label{w3}
\end{figure}

\begin{figure}[t!]
    \includegraphics[width=8cm]{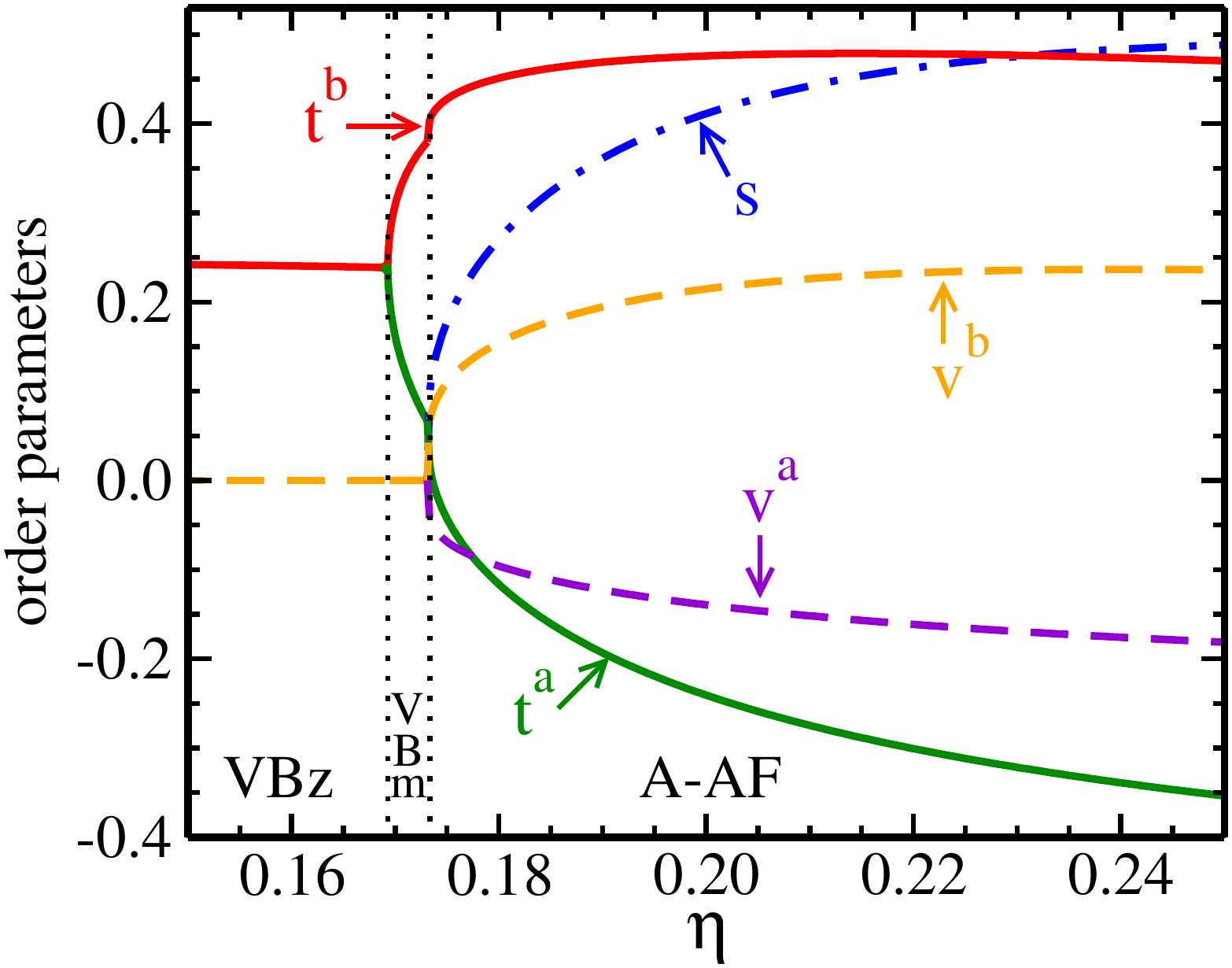}
    \caption{(Color online) Order parameters $\{s,t^{a,b},
                  v^{a,b}\}$ for $E_z =-0.72J$ and $0.15<E_z/J<0.25$ in the
                  VB$z$, VBm and $A$-AF phases, from left to right.}
    \label{w4}
\end{figure}

The ground state is characterized by order parameters obtained
directly during the self--consistency iterations in each phase:
spin, orbital and spin-orbital order parameters,
$\{s_1,t^{a,b}_{m,1},u^{a,b}_{m,1}\}$. We focus here on the phases
shown in the phase diagram of Fig. \ref{diag}. For the physical
reasons it is however better justified to define joint
spin-orbital order parameter in a slightly different way,
introducing a new variable $v^{\gamma}_{m,i}$ as follows:
\begin{eqnarray}
v^{\gamma}_{m,i}\equiv\langle S^z_i\tau^{\gamma}_i\rangle ,
\label{vdef}
\end{eqnarray}
which differs from the old order parameter by a subtraction of the
spin field, i.e., $u^{\gamma}_{m,i}=\frac12 s_i-v^{\gamma}_{m,i}$.
Now one can study the behavior of order parameters along different
cuts of the phase diagram of Fig. \ref{diag} and determine types
of phase transitions. Below we present a few representative
results. For this purpose we first choose $\eta=0.05$ and start
within the VB$z$ phase, where by increasing $E_z$ one gets first
into the PVB and next to $G$-AF phase, see Fig. \ref{w1}. For
$\eta=0.15$ there are even more phases and one passes through the
$A$-AF, ESO, EPVB, PVB, and AF-PVB phases, before reaching finally
the $G$-AF phase, see Figs. \ref{w2} and \ref{w3}. We also
investigated the dependence of order parameters on Hund's exchange
coupling --- we fixed $E_z=-0.72J$, started in the VB$z$ phase and
increased $\eta$ to get to the VBm and $A$-AF phases --- these
results are shown in Figs. \ref{w4}.

In what follows we use shorthand notation for the order parameters,
\begin{eqnarray}
\{s,t^{a,b},v^{a,b}\}\equiv\{s_1,t^{a,b}_{m,1},v^{a,b}_{m,1}\}.
\end{eqnarray}
In Fig. \ref{w1} we displayed the order parameters
for increasing $E_z$ in phases VB$z$, PVB and $G$-AF (separated by dotted
lines in the plot). The sublattice magnetization $s$ is non--zero only in
the $G$-AF phase because the remaining phases are of the VB crystal type,
with spin singlets oriented either along the $c$ direction or in the
$ab$ planes. In the $G$-AF phase the spin order grows stronger for
increasing $E_z$ when the orbital fluctuations weaken and spin
fluctuations present in the $G$-AF phase reduce $s$
from the classical value of 1/2.

Consider now decreasing values of $E_z$ in Fig. \ref{w1}. Both
orbital order parameters remain equal and close to $-1/4$ in the
$G$-AF phase until the (first order) transition point to the PVB
phase, where orbital configuration changes abruptly and becomes
anisotropic. In this case the $a$--$b$ symmetry was broken in such
a way that that spin singlets point in the PVB phase in $b$
direction and so the directional orbitals (cigars) do. This
explains the robust orbital order with $t^b$ being close to $-1/2$
in most of the PVB phase. The global symmetry is not broken as the
VB singlets form here a checkerboard pattern in the $ab$ plane,
with AO order of directional orbitals along the $a$ and $b$ axis
in the neighboring plaquettes. The transition to the VB$z$ phase
is discontinuous (first order) in the orbital sector too: $t^a$
grows constantly while decreasing $E_z$ down to $0.4J$, drops
slightly close to the transition point and jumps to $1/4$ in the
VB$z$ phase, $t^b$ grows quickly to $t^b\approx 0.125$ while
approaching the transition and then jumps to the value of $t^a$.
Qualitatively this means that close to the above transition the
orbital cigars pointing along the $b$ axis change gradually into a
shape very similar to clover orbitals lying in the $bc$ plane and
then suddenly the lobes along the $b$ direction disappear and we
are left with the pure VB$z$ phase.

The spin-orbital order parameter behaves in a much less intriguing
way; it remains zero in the VB$z$ and PVB phases, jumps to finite
value at the PVB-to-$G$-AF transition and remains almost constant
and close to $-0.1$ in the $G$-AF phase. The vanishing value of
$v^{a,b}$ in the singlet phases is simple to understand: the
orbitals are here fixed and spins form singlets and fluctuate
independently between the values $\pm 1/2$. This means that
$t^{a,b}$ and $s$ are not "synchronized" in any way and only this
could lead to $v^{a,b}\not=0$. This condition is satisfied in the
$G$-AF phase; orbitals are fixed and the spin configuration is
here determined by $s$ order parameter.

Figure \ref{w2} shows that the transition between the PVB and
$G$-AF phases can have a completely different character than
described above. The difference comes from a higher value of
$\eta$ which is now equal to $0.15$, enhancing the FM channel of
superexchange and leading to the intermediate PVB-AF phase and to
a smooth transition from the PVB to $G$-AF phase. In the PVB-AF
phase staggered magnetization $s$ grows continuously from zero (in
the PVB) to a finite value in the $G$-AF phase and remains
saturated there. This means that planar singlets in the PVB phase
decay gradually and spins get partially "synchronized" with
orbitals, moving toward uniform configuration which gives finite
spin-orbital order parameters $v^{\gamma}\neq 0$. The anisotropy
$(v^a\neq v^b)$ follows from the anisotropy of orbitals inherited
from the PVB order. This mechanism of the PVB-to-$G$-AF transition
is absent for low values of $\eta$ --- we anticipate that the
enhanced FM component of interactions reduces spin fluctuations
which makes the correlations between spins and orbitals possible.

In Fig. \ref{w3} we focus on the complementary regime of the phase
diagram, $\eta=0.15$ and negative $E_z$. In this regime we
describe three different consecutive phase transitions between the
phases: $A$-AF, ESO, EPVB and PVB. The first phase transition can
be regarded as a little bit artificial because this is a meeting
point of two completely different types of spin and orbital order,
with different symmetries and sublattices. For this reason the
transition has to be discontinuous and the spin order parameter
has different physical meaning on both sides of the transition
line, i.e., $s$ in the magnetic moment in the $A$-AF phase while
it is a weak AF order parameter in the ESO phase. We anticipate
that a smooth crossover occurs in place of such a transition in
the thermodynamic limit, nevertheless by comparing the energies we
concluded that this transition follows from the cluster MF
approach. Note also that the ESO phase has predominantly $z$
orbitals accompanied by fluctuations, i.e., $t^c\simeq -0.4$ and
$t^a=t^b$, and may be seen as an extension of the VB$z$ phase.

On the contrary, the second quantum EPVB phase which occurs in the
phase diagram of Fig. \ref{diag} may be seen as a precursor of the
PVB phase and is characterized again by finite joint spin-orbital
fluctuations, with $v^{\alpha}\neq 0$ for $\alpha=a,b$. What is
especially peculiar in the EPVB phase is the non--zero staggered
magnetization $s$ which grows smoothly from the zero values at the
phase borders meaning that we have a wedge of antiferromagnetism
between two VB configurations. The EPVB phase seems to be similar
to PVB-AF in a sense that spin-orbital fields are non--zero and
non--uniform but the qualitative behavior of the order parameters
is different, e.g. in the EPVB phase spin-orbital fields have
always opposite signs, while in the PVB-AF phase the sings are the
same.

Looking at the orbital order parameters $t^{a,b}$ in the $A$-AF
phase (Fig. \ref{w3}), one observes similar anisotropy as in the
PVB one but this time $a$-$b$ symmetry is not broken within the
cluster because in the $A$-AF phase every orbital is rotated by
$\pi/2$ with respect to its neighbors in the $ab$ plane. Another
difference is that the orbitals take the shape of clovers, not
cigars, with symmetry axes pointing along the $a$ or $b$ axis
which is described by $t^b$ being close to $1/2$. In the $A$-AF
phase we have also long--range magnetic order and finite
spin-orbital fields, indicating joint behavior of spin and orbital
MF variables.

Next Fig. \ref{w4} shows the behavior of order parameters for
$E_z=-0.72J$ and $\eta$ changing in an interval allowing us to
study the transitions from the VB$z$ to VBm phase, and between the
VBm and $A$-AF phase. In this case all the phases can be described
by the same spin and orbital sublattices because VB$z$ is uniform
in the orbital sector and has no long--range magnetic order so it
can be described both in terms of the PVB and $A$-AF type of
ordering. Global magnetization appears only in the $A$-AF phase
jumping from the zero value in the VBm and growing with increasing
$\eta$. Transition from the VB$z$ to VBm phase is continuous in
both spin and orbital sectors.

The orbital order parameters $t^{a,b}$ bifurcate in Fig. \ref{w4}
at $\eta\simeq 0.169$ from the isotropic value $t^a=t^b\simeq 1/4$
and the orbital anisotropy grows in the VBm phase to give AO order
in the $A$-AF phase (Fig. \ref{w4}), and next shows a
discontinuity at the second transition. The final AO order can be
described by clover orbitals with symmetry axes alternating
between $a$ and $b$ directions from site to site. Relatively big,
negative value of $t^a$ means that the clovers' lobes are
elongated in the $a$ or $b$ direction, perpendicular to their
axes. The elongation depends also on the value if $E_z$: the
$E_z\to -\infty$ limit corresponds to pure clover--like orbitals,
while for $E_z\to\infty$ one gets pure cigars. This tendency is
especially visible in the FM phase which is not limited in
horizontal direction of the phase diagram. Consequently, the VBm
phase can be regarded as a crossover regime between orbitally
uniform VB$z$ and alternating $A$-AF phases. This resembles to
some extent the PVB-AF phase described earlier but we want to
emphasize the main difference between these phases: the VBm phase
does not need non--factorizable spin-orbital MF to appear while
the PVB-AF one needs it (compare Figs. \ref{diagns} and
\ref{diag}). The question of spin-orbital non--factorizability
will be addressed in more details below, see Sec. \ref{sec:enta}.

\begin{figure}[t!]
    \includegraphics[width=8cm]{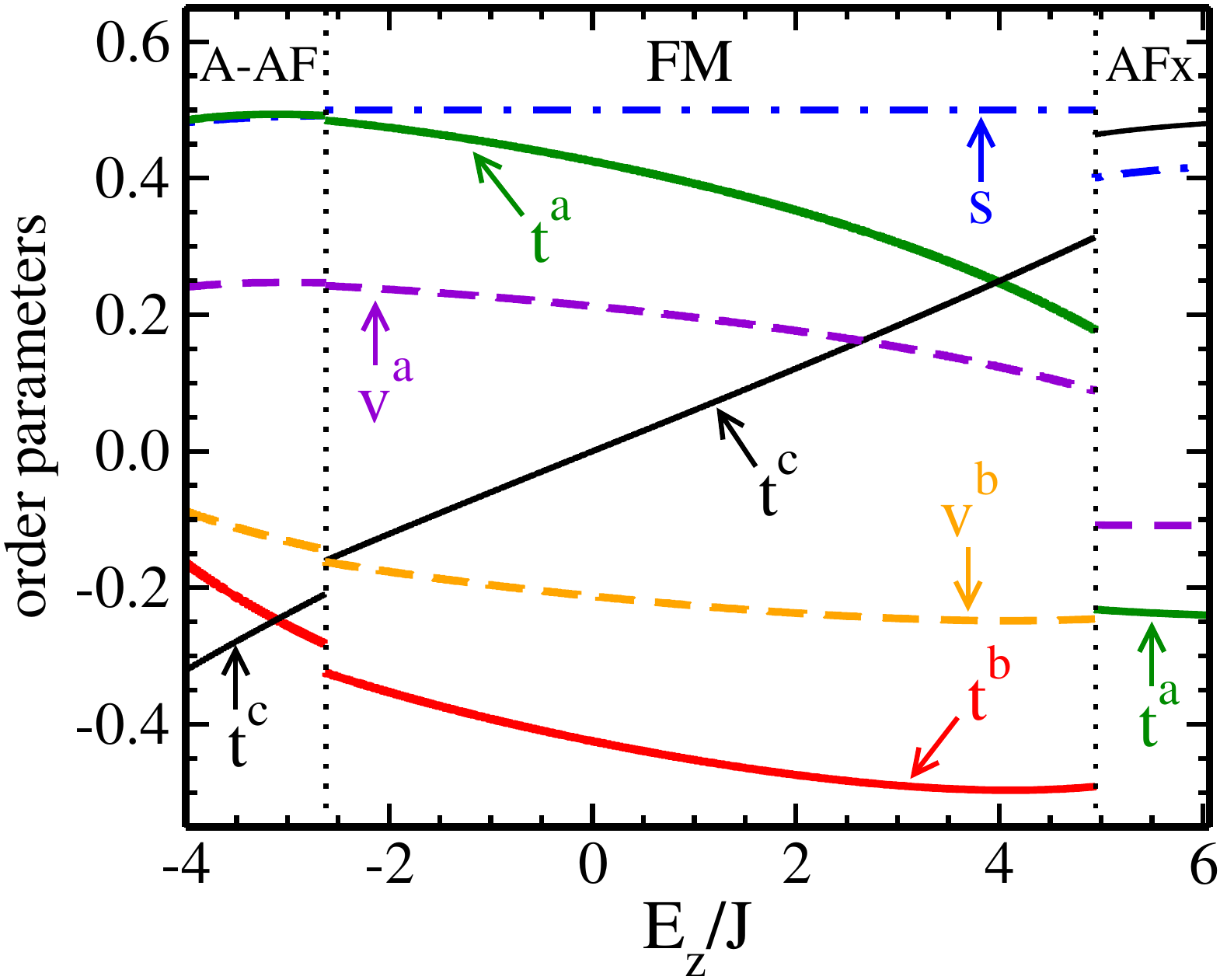}
    \caption{(Color online) Order parameters $\{s,t^{a,b,c},v^{a,b}\}$
                   for $\eta =0.28$ and $-4.0<E_z/J<6.0$ in the
                   $A$-AF, FM and $G$-AF phases, from left to right.}
    \label{w5}
\end{figure}

Finally, we show the behavior of the order parameters
$\{s,t^{a,b,c},v^{a,b}\}$ and the quantum fluctuation effects on
them in the $A$-AF, FM and $G$-AF phases for $\eta=0.28$, see Fig.
\ref{w5}. The third orbital field $t^c$ is linearly dependent on
$t^a$ and $t^b$ (by the constraint $t^c=-t^a-t^b$), and was added
here to visualize the orbital order along the $c$ direction which
is essential in the large $|E_z|$ regime showed in Fig. \ref{w5}.
In FM phase the spin order is saturated because of the lack of
quantum and thermal fluctuation. For the same reason spin-orbital
field factorizes and $\{v^a,v^b\}$ fields bring no extra
information which would not be already contained in $\{t^a,t^b\}$.
The overall behavior of $t^c$ is in agreement with the crystal
field part of the Hamiltonian Eq. (\ref{hamik}) with
$t^c\to\pm1/2$, giving uniform cigar or clover orbitals depending
on the sign of $E_z$.

We emphasize that for increasing $E_z$ one finds two crossing
points of $t^c$ with $\{t^a,t^b\}$ curves, one at $t^c=t^b=-1/4$
and the other one at $t^c=t^a=1/4$. At these two points the
orbitals take shapes of perfect clovers ($E_z<0$) or perfect
cigars ($E_z>0$), with symmetry axes alternating in the $ab$ plane
from site to site. Only one of these points belongs to the FM
phase meaning that the four "perfect" orbital configurations: AO
order with clovers/cigars and FO order with clovers/cigars are
separated by phase transitions in the spin-orbital model Eq.
(\ref{hamik}). The transitions shown in Fig. \ref{w5} are
discontinuous due to the change of global spin order in each
phase. The spin order parameter $s$ plays a role of staggered
$A$-AF or AF magnetization in the extremal phases and is trivial
(saturated) in the FM phase. On the other hand, all three phases
displayed in Fig. \ref{w5} can be described by the same orbital
sublattices assuming AO order. The large scale of $E_z$ in Fig.
\ref{w5} is in contrast to those in other figures
--- it indicates that orbital degrees of freedom are very rigid
when spins are almost frozen and one needs rather high energies to
change their configuration.

\section{Nearest--neighbor correlations}
\label{sec:corr}

\subsection{Spin, orbital and spin-orbital correlations}
\label{sec:nn}

Studying order parameters in different phases we get complete
information about symmetry broken or disordered phases of the
system, but this alone does not justify the use of the cluster MF
method as order parameters can in principle be obtained using
standard single-site MF approximation, see Sec. \ref{sec:ssa}. The
advantage of the cluster method becomes evident when we
investigate correlation functions on the bonds belonging to the
considered cube. The most obvious ones are the spin--spin
correlations $\langle{\bf S}_i\cdot{\bf S}_j\rangle$ or
orbital--orbital correlations
$\langle\tau^{\gamma}_i\tau^{\gamma}_j\rangle$, but in addition
one may also determine joint spin-orbital correlations,
$\langle({\bf S}_i\cdot{\bf S}_j)\tau^{\gamma}_i
\tau^{\gamma}_j\rangle$. Although one could in principle invent
several other bond correlation functions, the above ones have the
most transparent physical meaning because they enter the
Hamiltonian. For the same reason we will only consider orbital
correlation functions for different bond direction $\gamma=a,b,c$.
This gives nine correlation functions, three in each direction,
for each vertex of the cube. For symmetry reasons it is enough to
consider only one chosen vertex, e.g. vertex number 1 in Fig.
\ref{cub}. For convenience we will use the following notation:
\begin{eqnarray}
C_s^{\gamma}&\equiv&\langle {\bf S}_1\cdot{\bf S}_i \rangle , \\
C_t^{\gamma}&\equiv&\langle \tau^{\gamma}_1\tau^{\gamma}_i \rangle , \\
C_{st}^{\gamma}&\equiv&
\langle ({\bf S}_1\cdot{\bf S}_i)\tau^{\gamma}_1\tau^{\gamma}_i\rangle ,
\end{eqnarray}
where the bond $\langle 1i\rangle||\gamma$ and $i\in\{2,3,7\}$
which gives all nonequivalent nearest neighbor correlations along
$\gamma=a,b,c$ (see Fig. \ref{cub}).

In the next paragraphs we will present the numerical results for
bond correlations $\{C_s^{\gamma},C_t^{\gamma},C_{st}^{\gamma}\}$
along different cuts of the phase diagram of Fig. \ref{diag}. For
all three--panel plots each panel describes correlation of one
type: upper panel --- spin correlations, middle one
--- orbital correlations, and bottom one --- spin-orbital
correlations. For each panel different characters (colors) of line
indicate different direction $\gamma$: solid (red) line stands for
$\gamma=c$, dashed (green) line for $\gamma=a$ and dashed--dotted
(blue) line for $\gamma=b$. In case of two--panel plots there are
only two directions considered, $c$ and $a$ because for symmetry
reasons correlations along the $b$ and $a$ axes are identical.
Therefore, the left panel concerns all three types of correlators
for $\gamma =c$ and the right one for $\gamma =a$ in the way that
solid (red) lines are spin--spin correlation functions, dashed
(green) ones are orbital--orbital correlations and dashed--dotted
(blue) represent spin-orbital correlators. In order to investigate
the nature of spin-orbital entanglement we focus the discussion on
two quantum phases which occur at finite values of Hund's exchange
$\eta$ near the orbital degeneracy: (i) the PVB phase, and (ii)
the ESO phase.

\subsection{Plaquette valence-bond phase}
\label{sec:pvb}

\begin{figure}[t!]
    \includegraphics[width=8.0cm]{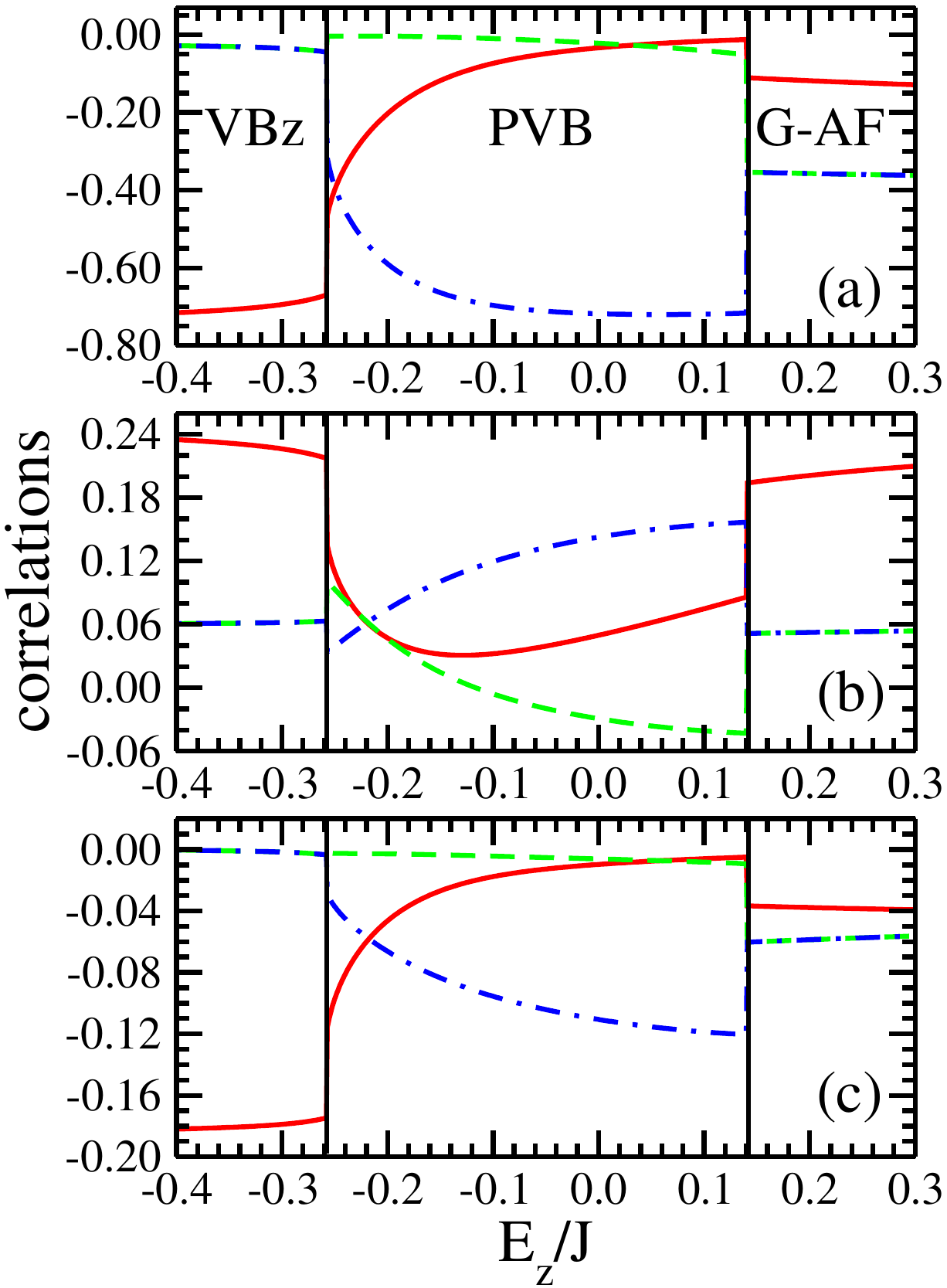}
    \caption{(Color online) Nearest neighbor correlations
                  for $\eta =0.05$ and $-0.4<E_z/J<0.3$ in the
                  VB$z$, PVB and $G$-AF phases. (a) Spin correlations:
                  solid (red) line --- $C_s^{c}$, dashed (green)
                  line --- $C_s^{a}$ and
                  dashed--dotted (blue) line --- $C_s^{b}$.
                  (b) Orbital correlations: solid (red)
                  line --- $C_t^{c}$, dashed (green)
                  line --- $C_t^{a}$ and
                  dashed--dotted (blue) line --- $C_t^{b}$.
                  (c) Spin-orbital correlations: solid (red)
                  line --- $C_{st}^{c}$, dashed (green)
                  line --- $C_{st}^{a}$ and
                  dashed--dotted (blue) line --- $C_{st}^{b}$.
                  }
    \label{c1}
\end{figure}

We begin with bond correlation functions for $\eta=0.05$ and
$-0.4<E_z/J<0.3$ in the VB$z$, PVB and $G$-AF phase. The $C_s^{c}$
function stays close to $-3/4$ in the VB$z$ phase while the other spin
correlations are almost zero as one can expect in the interlayer
singlet phase, see Fig. \ref{c1}(a). After the first transition at
$E_z\simeq -0.26J$ the situation changes --- now the singlets are in
$b$ direction and $C_s^{b}$ gets close to $-3/4$ when $E_z$
increases. After the second transition at $E_z\simeq 0.14J$ all
the spin correlations take finite negative values with $C_s^{c}$
relatively weakest, keeping the symmetry between $a$ and $b$
direction. This is in agreement with the spin order in the $G$-AF
phase discussed in Sec. \ref{sec:ops}.

\begin{figure}[t!]
    \includegraphics[width=8.0cm]{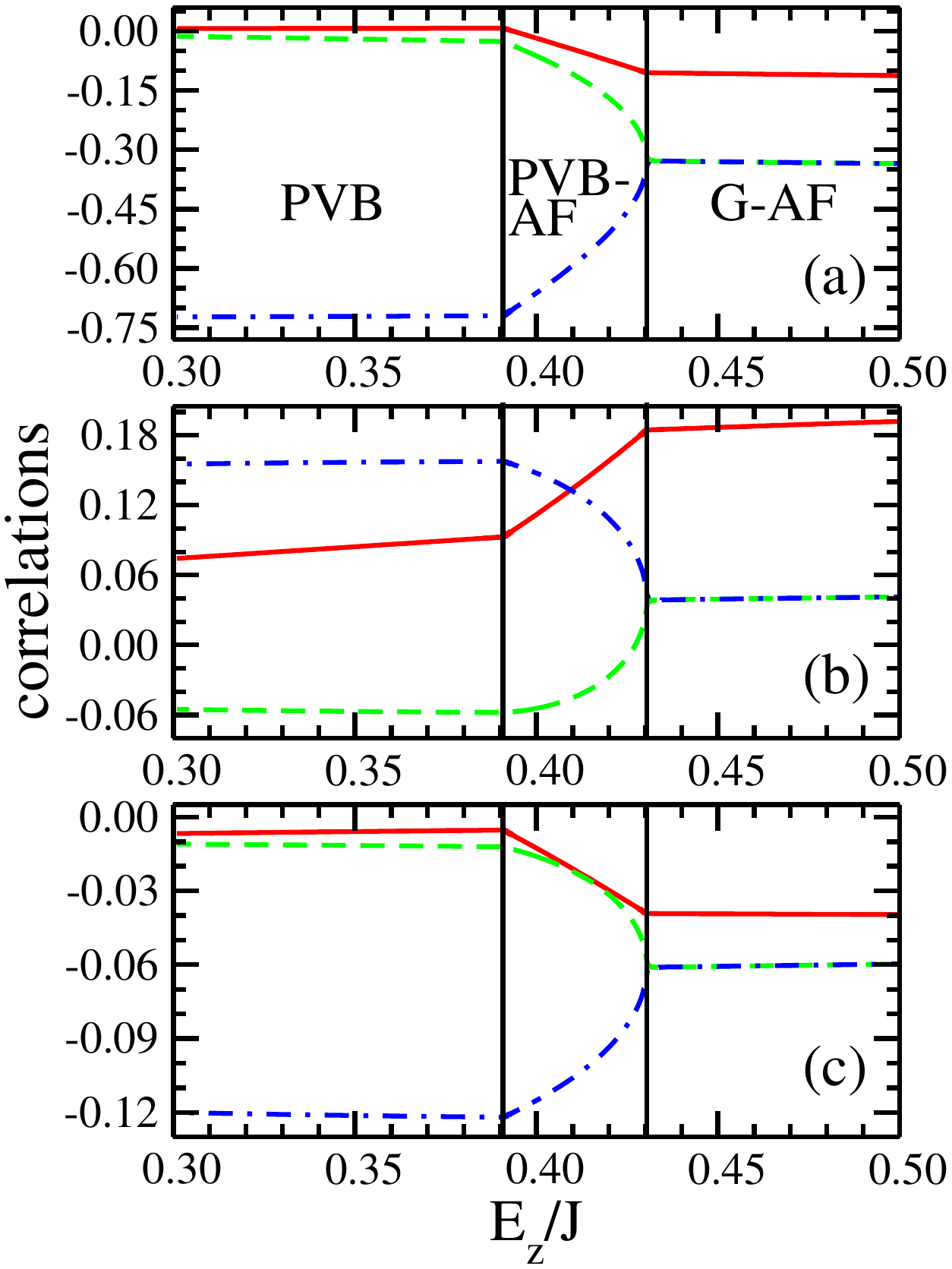}
    \caption{(Color online) Nearest neighbor correlations
                  for $\eta =0.15$ and $0.3<E_z/J<0.5$ in the
                  PVB, PVB-AF and $G$-AF phases. (a) Spin correlations:
                  solid (red) line --- $C_s^{c}$, dashed (green)
                  line --- $C_s^{a}$ and
                  dashed--dotted (blue) line --- $C_s^{b}$.
                  (b) Orbital correlations: solid (red)
                  line --- $C_t^{c}$, dashed (green)
                  line --- $C_t^{a}$ and
                  dashed--dotted (blue) line --- $C_t^{b}$.
                  (c) Spin-orbital correlations: solid (red)
                  line --- $C_{st}^{c}$, dashed (green)
                  line --- $C_{st}^{a}$ and
                  dashed--dotted (blue) line --- $C_{st}^{b}$.}
    \label{c2}
\end{figure}

The orbital correlation functions in the VB$z$ and $G$-AF phases
behave as if the orbitals were frozen in uniform configuration
with $t^c=\pm1/2$ and $t^{a,b}=\mp1/4$ whereas in the PVB phase
their behavior is more nontrivial; the dominant $C_t^{b}$ is quite
distant from its maximal value $1/4$ and the difference between
$C_t^a$ and $C_t^c$ is visible, especially close to the $G$-AF
phase, see Fig. \ref{c1}(b). This result is due to quantum
fluctuations: perfect VB$z$ and $G$-AF configurations are the
exact eigenstates of the Hamiltonian, at least in the limit of
large $|E_z|$, while perfect PVB state cannot be obtained exactly
in any limit and gets easily destabilized by varying $E_z$. It is
peculiar that the spin configuration is almost nonsensitive to the
orbital splitting $E_z$ and the singlets stay rigid in the regime
of spin disordered phases, i.e., below the transition to the
$G$-AF phase. The spin-orbital sector, shown in Fig. \ref{c1}(c),
does not bring any new information; all the lines behave as if
spin and orbital degrees of freedom were factorizable.

Figure \ref{c2} presents the bond correlations for a gradual
transition between the PVB and $G$-AF phases, with an intermediate
PVB-AF phase for $\eta =0.15$ and $0.3<E_z/J<0.5$. By decreasing
$E_z$, i.e., looking from right to left, we can see the in--plane
spin correlation bifurcating smoothly at the transition to the
PVB-AF phase and evolving monotonically to the values
characteristic of the PVB phase, see Fig. \ref{c2}(a). The
interplane spin correlations $C_s^{c}$ stay relatively weak
everywhere which is obvious in both PVB and $G$-AF phase and hence
not so surprising in the intermediate PVB-AF phase.

In the orbital sector we can see here very similar behavior to the
one observed in Fig. \ref{c1} --- again the order is far from the
perfect PVB but $C_t^{c}$ is close to the classical value of
$1/16$ obtained for the plane perpendicular to two directional
orbitals along the $b$ axis, while $C_t^{a}$ is almost exactly
opposite and $C_t^{b}$ stays below $1/4$, see Fig. \ref{c2}(b).
This shows some kind of universality at the transition from the
PVB to $G$-AF phase which is independent of the intermediate
phase. Again, the spin-orbital sectors, shown in Fig. \ref{c2}(c),
does not indicate any qualitatively new behavior comparing to
spins and orbitals alone but looking at the phase diagrams with
(Fig. \ref{diagns}) and without (Fig. \ref{diag}) spin-orbital
factorization we recognize that on-site spin-orbital entanglement
must be responsible for the onset of the PVB-AF phase.

\subsection{Phases with entangled spin-orbital order}
\label{sec:eso}

\begin{figure}[t!]
    \includegraphics[width=7.7cm]{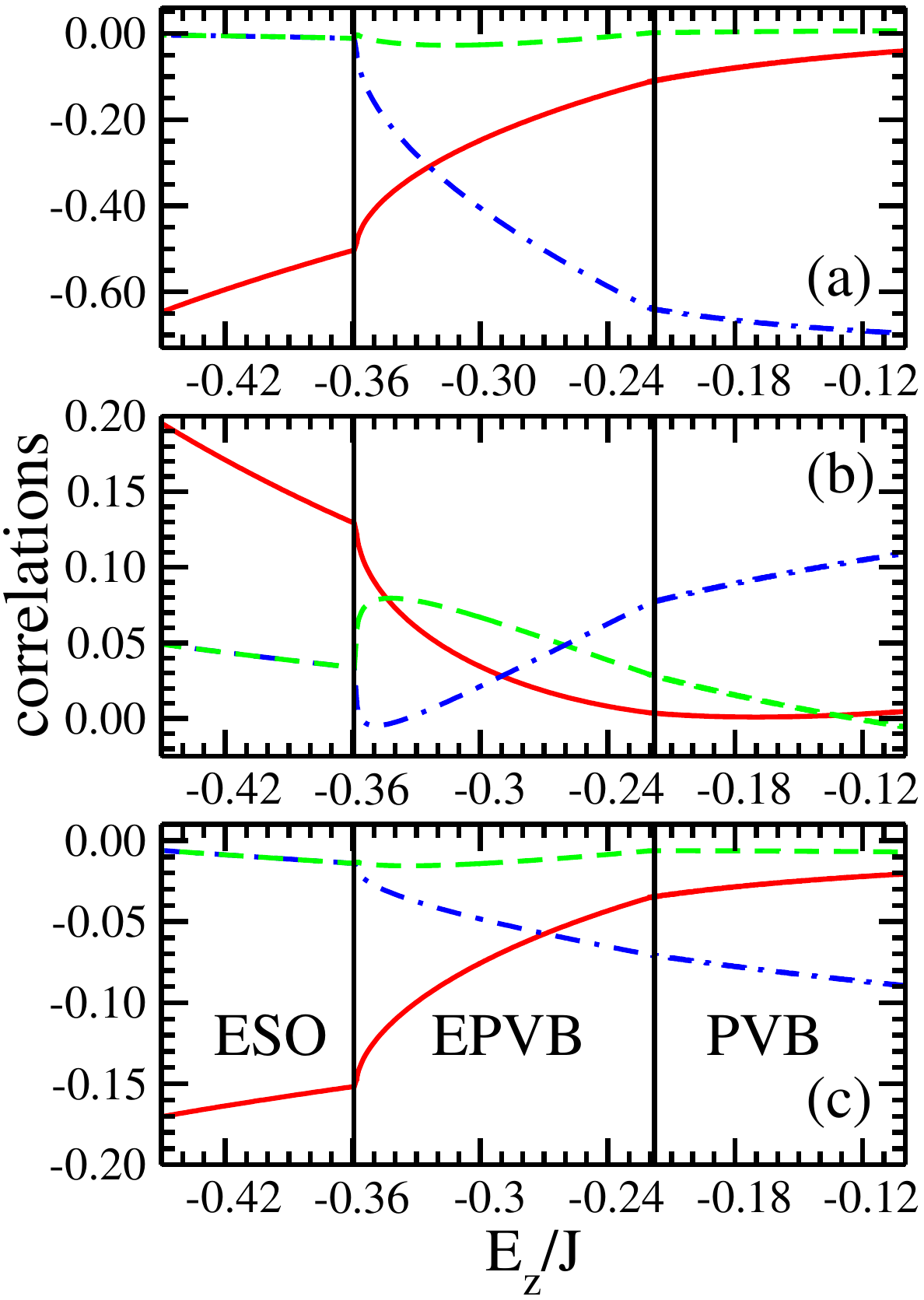}
    \caption{(Color online) Nearest neighbor correlations
                  for $\eta=0.15$ and $-0.45<E_z/J<-0.1$ in the
                  ESO, EPVB and PVB phases. (a) Spin correlations:
                  solid (red) line --- $C_s^{c}$, dashed (green)
                  line --- $C_s^{a}$ and
                  dashed--dotted (blue) line --- $C_s^{b}$.
                  (b) Orbital correlations: solid (red)
                  line --- $C_t^{c}$, dashed (green)
                  line --- $C_t^{a}$ and
                  dashed--dotted (blue) line --- $C_t^{b}$.
                  (c) Spin-orbital correlations: solid (red)
                  line --- $C_{st}^{c}$, dashed (green)
                  line --- $C_{st}^{a}$ and
                  dashed--dotted (blue) line --- $C_{st}^{b}$.}
    \label{c3}
\end{figure}

Consider now smaller (negative) values of $E_z$, where unexpected
and qualitatively new entangled phases occur in the phase diagram
of Fig. \ref{diag}. We display bond correlation functions in Fig.
\ref{c3} in two neighboring highly frustrated and entangled
phases, the ESO and EPVB phase --- the latter one turns into the
PVB phase when $E_z$ is increased. The relevant parameter range
for $\eta =0.15$ is $-0.45<E_z/J<-0.1$. On the first glance this
plot shows that the transitions between the ESO and EPVB as well
as between the EPVB and PVB phases are of the second order. In the
spin sector one observes weakening singlet order in the ESO phase
with $C_s^{c}$ getting far from $-3/4$ and in-plane correlations
$C_s^{a,b}$ being practically vanishing, see Fig. \ref{c3}(a).
After the first transition (at $E_z\simeq -0.36J$) $C_s^b$ grows
rapidly toward negative values while $C_s^c$ goes to zero much
more gently and $C_s^a$ stays close to zero. This means that in
the EPVB phase we have relatively strong AF order in the $bc$
plane inside the cluster, turning into the $ac$ plane order on
neighboring cubes. This gives finite magnetization $s$ shown in
Fig. \ref{w3}. When approaching the second transition (at
$E_z\simeq -0.22J$) $C_s^c$ weakens and $C_s^b$ gets closer to
$-3/4$ and this is continued within the PVB phase.

In the orbital sector we can find other differences between
entangled and disentangled phases, see Fig. \ref{c3}(b). In the
ESO phase the $C_t^c$ drops considerably when approaching the
first transition; this is in contrast with the VB$z$ phase where
$C_t^c$ stays almost constant until the transition occurs.
However, one finds that the spin-orbital bond correlation
$C_{st}^c$ stays constant in the ESO phase, see Fig. \ref{c3}(c).
The behavior of in-plane correlation functions $C_t^{a,b}$ becomes
somewhat puzzling within the EPVB phase: after bifurcation at the
transition point $C_t^b$ drops to zero and slowly recovers to
become dominant in the PVB phase, while $C_t^a$ stays dominant in
certain region of the EPVB phase even though the spin correlations
in $a$ direction vanish. Only $C_t^c$ gradually drops to zero
throughout all three phases.

\begin{figure}[t!]
    \includegraphics[width=8.2cm]{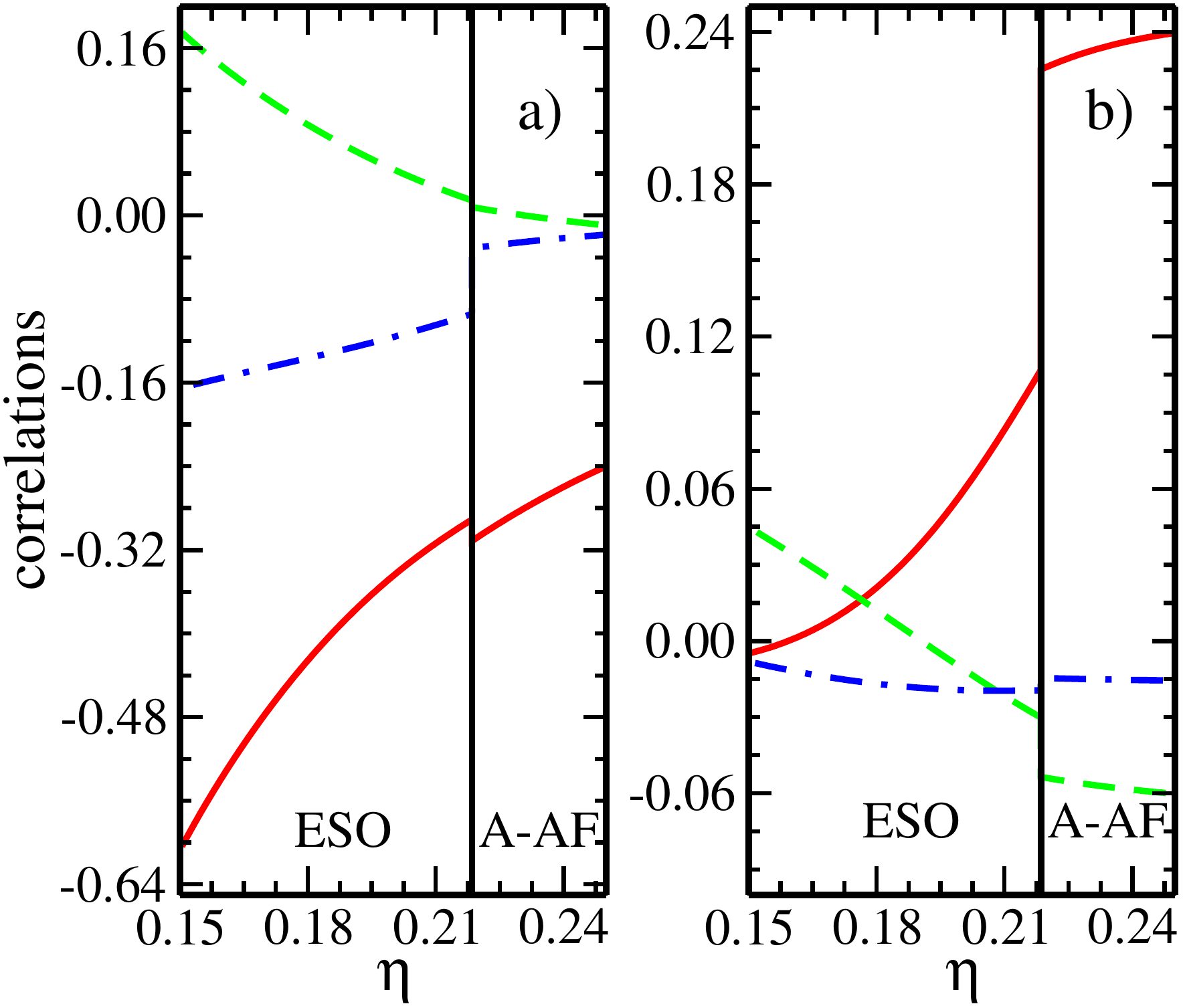}
    \caption{(Color online) Nearest neighbor correlations
                  for $E_z=-0.43J$ and $0.15<\eta<0.25$ in the
                  ESO and $A$-AF phases. (a) Correlations along the $c$ axis:
                  solid (red) line --- $ C_s^{c}$, dashed (green)
                  line --- $ C_t^{c}$ and
                  dashed--dotted (blue) line ---
                  $ C_{st}^{c}$.
                  (b) Correlations within $ab$ planes: solid (red)
                  line --- $ C_s^{a}$ , dashed (green)
                  line --- $ C_t^{a}$ and
                  dashed--dotted (blue) line ---  $ C_{st}^{a}$.
                  Correlations in $a$ and $b$ direction are the same.}
    \label{c4}
\end{figure}

Note that in the spin-orbital sector we can see the joint order in
both entangled phases in a more transparent way than in the
orbital one, at least concerning the ESO and EPVB phases (we
should keep in mind that $-3/16\leq C_{st}^{\gamma}\leq 1/16$
while $-1/4\leq C_{t}^{\gamma}\leq 1/4$ where the bottom limit for
$C_{st}^{\gamma}$ is realized only in singlet phases). The
$C_{st}^c$ correlation is definitely dominant in the ESO phase and
stays dominant in most of the EPVB phase in contrary to spin
$C_{s}^c$ correlation. In addition, close to the second transition
the $C_{st}^c$ correlation is overcome by $C_{st}^b$ which grows
here stronger because of singlets being formed on the bonds along
the $b$ axis. This tendency is further amplified within the PVB
phase. Note that $C_{st}^a$ stays practically zero in all the
phases shown in Fig. \ref{c3}.

Now we turn to the dependence of bond correlations on increasing
Hund's exchange $\eta$. In Fig. \ref{c4} we display correlations
for $E_z=-0.43J$ and $0.15<\eta<0.25$ in the ESO and $A$-AF
phases. Both phases can be described by a strong tendency toward
AO order with two sublattices which does not violate the $a$--$b$
symmetry inside the cube; for this reason we show only
correlations along the $c$ and $a$ direction. The spin sector
within the ESO phase is dominated by the decay of interplanar
singlets accompanied by growth of in-plane correlations which
triggers global $A$-AF order above the transition (at $\eta\simeq
0.22$). The orbital correlations in the $c$ direction drop almost
to zero when $\eta$ grows and stay small in the $A$-AF phase. The
in-plane orbital correlations $C_t^{a,b}$ decrease in the ESO
phase too but remain finite after the transition. Summarizing, in
the ESO phase close to the onset of the $A$-AF one we find a very
weak orbital order accompanied by precursors of the $A$-AF order
in spin sector.

Consider now the spin-orbital correlations. In the ESO phase
$C_{st}^c$ takes relatively big, negative values and does not
change much except for the transition point where it jumps to
zero. In contrast, in the $A$-AF phase we no longer observe any
spin-orbital ordering. Note that a peculiar signature of the ESO
phase is rather robust spin-orbital order on the interlayer bonds
along the $c$ axis which turns out to be more rigid against
quantum fluctuations than orbital order and remains finite even
when orbital order vanishes.

\begin{figure}[t!]
    \includegraphics[width=8.2cm]{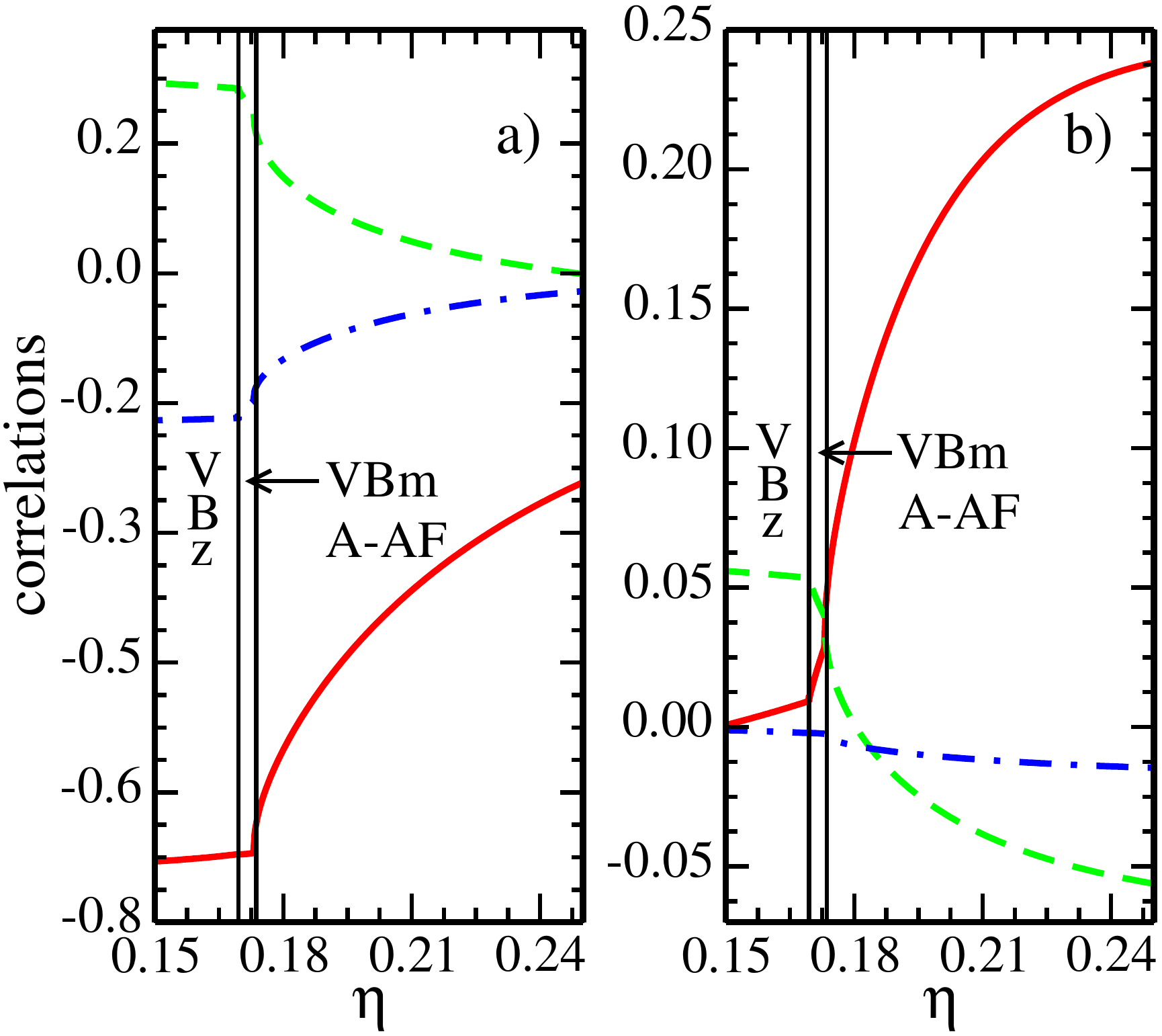}
    \caption{(Color online) Nearest neighbor correlations
                  for $E_z=-0.72J$ and $0.10<\eta<0.25$ in the
                  VB$z$, VBm and $A$-AF phases. (a) Correlations
                  along the $c$ axis: solid (red) line --- $C_s^{c}$,
                  dashed (green) line --- $C_t^{c}$ and
                  dashed--dotted (blue) line ---
                  $  C_{st}^{c}$.
                  (b) Correlations within $ab$ plane: solid (red)
                  line --- $C_s^{a}$ , dashed (green)
                  line --- $C_t^{a}$ and
                  dashed--dotted (blue) line ---  $ C_{st}^{a}$.
                  Correlations in $a$ and $b$ direction are the same.}
    \label{c5}
\end{figure}

In the last figure, Fig. \ref{c5}, we display bond correlation
functions in the VB$z$, VBm and $A$-AF phases for $E_z=-0.72J$ and
$0.10<\eta<0.25$. As before, all the in-plane correlations are
independent of $\gamma$.
The plots prove that the transition from the VB$z$ to VBm phase is
of the second order while the transition from the VBm to $A$-AF
phase produces no discontinuities in correlations either, but the
behavior of order parameters (see Fig. \ref{w4}) is here slightly
discontinuous. In the spin sector we observe first (at
$\eta<0.17$) that robust singlets along the $c$ axis with
$C_s^c\simeq -0.7$, see Fig. \ref{c5}(a), are gradually weakened
under increasing $\eta$ and weak FM correlations occur in the
VB$z$ phase close to the first phase transition to the VBm order.
We suggest that this regime of parameters could correspond to
K$_3$Cu$_2$F$_7$, where the magnetic properties indicate
interplanar singlets as formed in the VB$z$ and VBm phases
accompanied by weak FM correlations in the $ab$
planes.\cite{Man07}

Note that the changes in spin correlations with increasing $\eta$
become fast only after leaving the VBm phase. In the orbital
sector perfect VB$z$ order dies out quickly already in the VBm
regime, both on the bonds along the $c$ and $a$ axes. After
entering the $A$-AF phase, $C_t^{c}$ vanishes exponentially while
$C_t^{a}$ crosses zero and gradually falls to negative values.
This behavior is in agreement with that shown in Fig. \ref{w4}
saying that $t^c$ remains close to zero in the $A$-AF phase and
the negative $C_t^{a}$ confirms AO order in $ab$ planes.
Altogether, the spin-orbital sector does not exhibit here any
considerable non-factorizable features.

\section{Spin--orbital entanglement}
\label{sec:enta}

\begin{figure}[b!]
    \includegraphics[width=8.2cm]{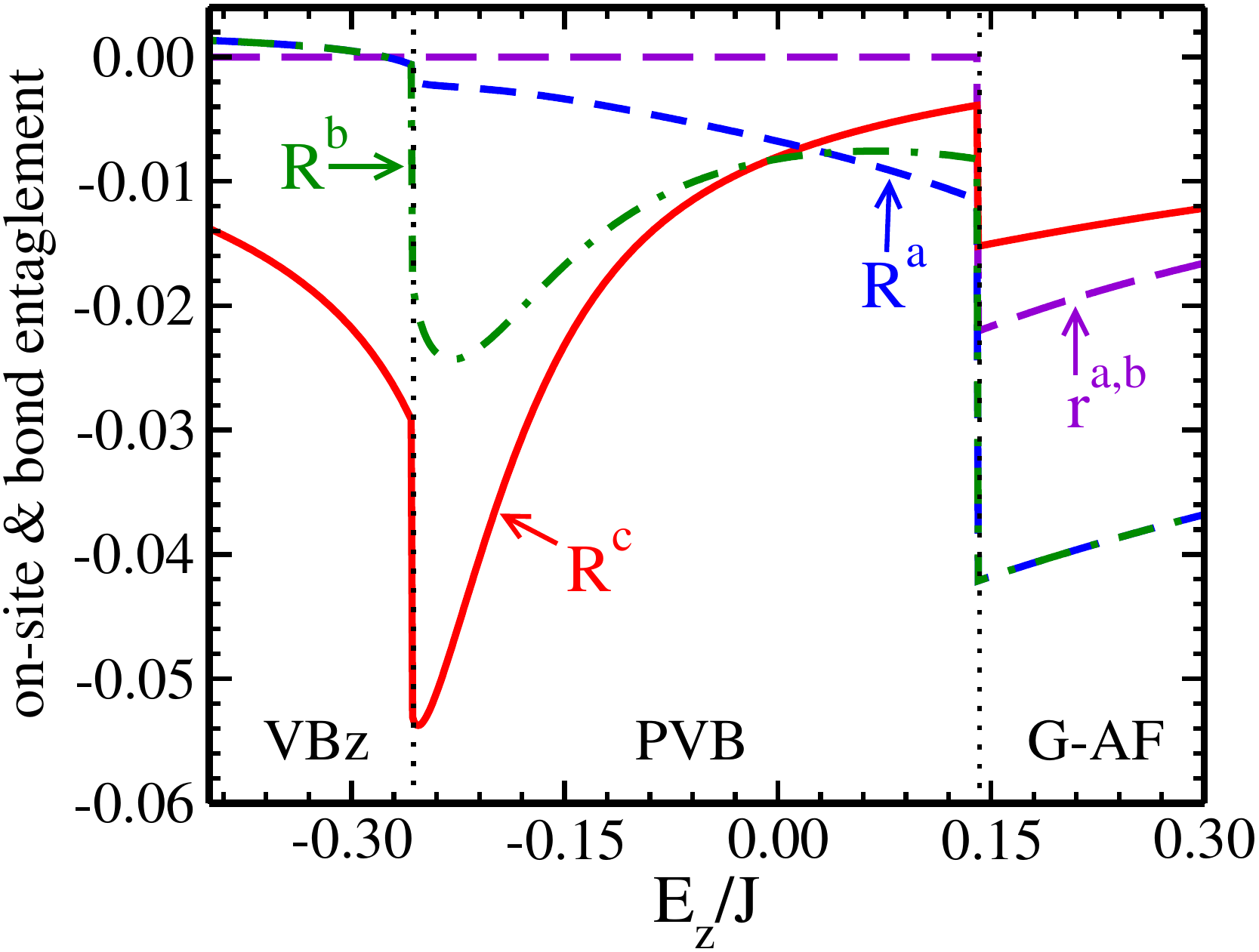}
\caption{(Color online) On-site $r^{\gamma}$ Eq. (\ref{entalo})
and bond $R^{\gamma}$ Eq. (\ref{entabo}) entanglement parameters
for $\eta =0.05$ and $-0.4<E_z<0.3$ in the VB$z$, PVB and $G$-AF
phases.}
    \label{e1}
\end{figure}

The essence of spin-orbital entanglement observed in the cluster
MF approach is spin-orbital non-factorizability. This feature can
have either on-site or bond character, the latter was introduced
in Ref. \onlinecite{Ole06}. We emphasize that on-site entanglement
which is characteristic for cases with finite spin-orbit
coupling,\cite{Jac09} occurs also in the present superexchange
model as shown below. We define the on-site entanglement as
non-separability of the order parameters, i.e., spin and orbital
operators are entangled when $v^{\gamma}\not=st^{\gamma}$, while
the entanglement as being of bond type when\cite{Ole06}
$C_{st}^{\gamma}\not=C_s^{\gamma}C_t^{\gamma}$, implying that it
can be detected by investigating the respective correlation
functions. Therefore we analyze in this Section the numerical
results for the quantities (covariances) motivated by the above
discussion which are defined as follows:
\begin{eqnarray}
\label{entalo}
r^{\gamma}&=&v^{\gamma}-st^{\gamma}\,, \\
\label{entabo}
R^{\gamma}&=&C_{st}^{\gamma}-C_s^{\gamma}C_t^{\gamma}\,.
\end{eqnarray}
In case of $r^{\gamma}$ we consider only $\gamma=a,b$ as the
on-site covariance satisfy the local constraint,
\begin{equation}
r^c=-r^a-r^b,
\end{equation}
while for $R^{\gamma}$ we shall present the data for
$\gamma=a,b,c$. In order to quantify the above non-factorizability
and to recognize whether it is strong or weak in a given phase, it
is necessary to establish first the minimal and maximal values of
$R^{\gamma}$ and $r^{\gamma}$. Simple algebraic considerations
give the following inequalities: the bond covariances
$|R^{\gamma}|<0.25$ in singlet phases, $|R^{\gamma}|<0.125$ in
phases with magnetic order, and the on-site covariances
$|r^{\gamma}|<0.25$ everywhere.

\begin{figure}[t!]
    \includegraphics[width=8.2cm]{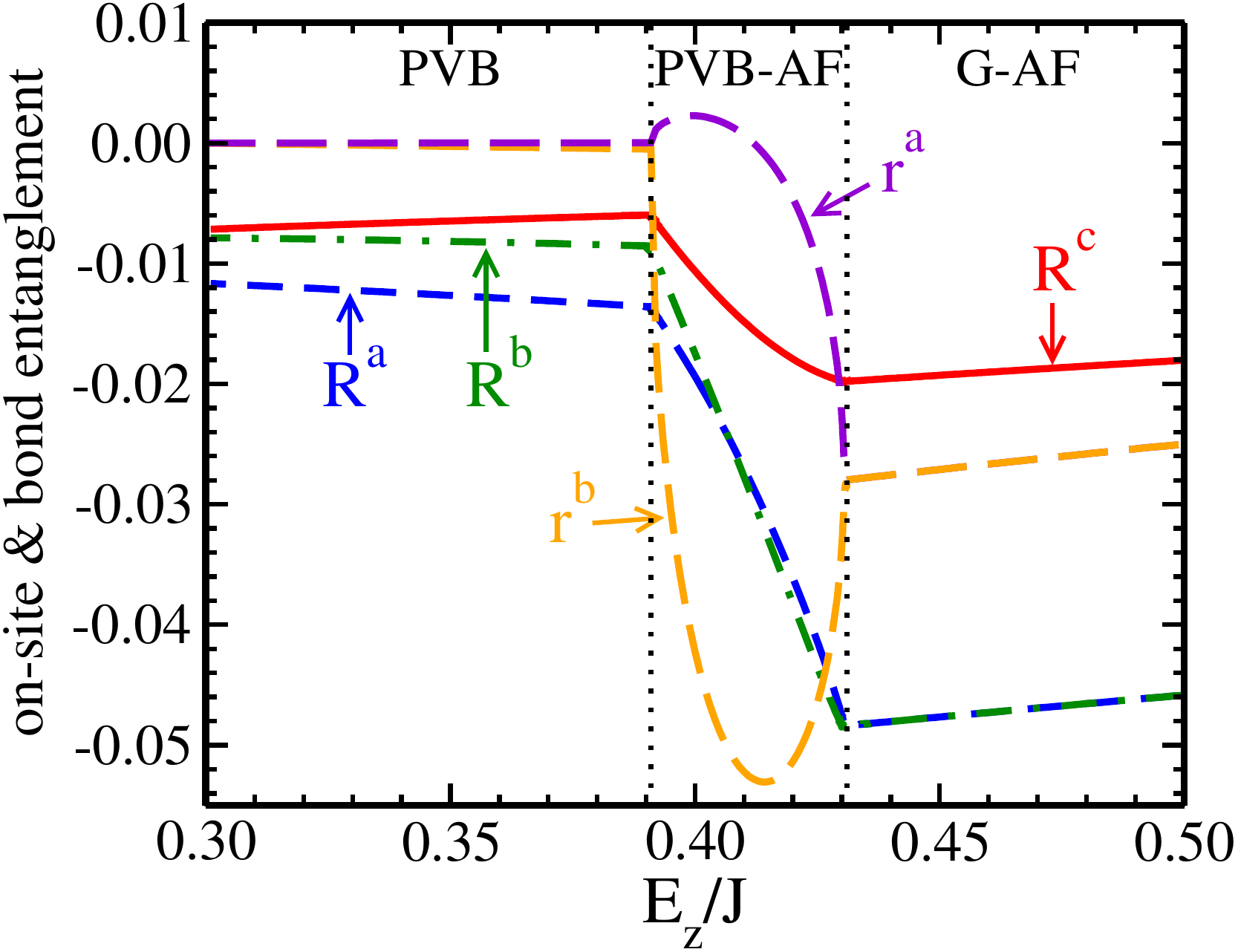}
    \caption{(Color online) On-site $r^{\gamma}$ Eq. (\ref{entalo})
and bond $R^{\gamma}$ Eq. (\ref{entabo}) entanglement parameters
for $\eta =0.15$ and $0.3<E_z/J<0.5$ in the
                  PVB, PVB-AF and $G$-AF phases.}
    \label{e2}
\end{figure}

First of all, the numerical results show that both bond Eq.
(\ref{entabo}) and on-site Eq. (\ref{entalo}) spin-orbital
entanglement is small in the regime of weak Hund's exchange
coupling. This feature is illustrated in Fig. \ref{e1} for the
VB$z$, PVB and $G$-AF phases at $\eta =0.05$ and $-0.4<E_z/J<0.3$.
The $r^a=r^b$ curves show no on-site spin-orbital entanglement
($r^{\gamma}=0$) in both VB$z$ and PVB phases, while it is finite
in the $G$-AF phase ($r^a=r^b<0$) and gradually approaches zero
with increasing $E_z$. We emphasize that this on-site
non-factorizability is minute, being one order of magnitude
smaller than its maximal value, and does not play any important
role for the stability of the $G$-AF ground state. This is
confirmed by the fact that $G$-AF phase exists in the same region
of parameters in both phase diagrams: factorizable (Fig.
\ref{diagns}) and non-factorizable one (Fig. \ref{diag}), and
occurs even in the single-site MF approximation (Fig.
\ref{ssmfa}). It is interesting to note that the in-plane bond
entanglement $R^{a,b}$ takes relatively high values in the $G$-AF
phase. This is clearly an effect of quantum fluctuations; the
perfect (classical) $G$-AF phase of Fig. \ref{ssmfa} has uniform
fixed $x$ orbital configuration with $t^c=1/2$ which suppresses
any non-factorizability. As the on-site entanglement, also the
bond spin-orbital entanglement vanishes gradually for high values
of $E_z\to\infty$.

At the border line between the VB$z$ and PVB phases we noticed a
considerable increase of $R^c$ and less pronounced growth of $R^b$
which seem to be induced by the transition as the $R^{b,c}$ drop
quickly for higher values of $E_z$. In the VB$z$ phase we expect
all the spin-orbital covariances to be zero for the same reasons
as in the $G$-AF phase and this also applies to the perfect PVB
phase. In Fig. \ref{e1}, however, the VB$z$ and PVB phases are
dominated by the critical behavior which distorts perfect
orderings.

Also in the regime of higher Hund's exchange interaction
$\eta=0.15$ the spin-orbital covariances in the PVB, PVB-AF and
$G$-AF phases are small in the range of their stability, see Fig.
\ref{e2} for $0.3<E_z/J<0.5$. In the PVB phase all the covariances
take small values showing that the PVB type of order has no
serious quantum fluctuations in this parameter range. The on-site
covariances $\{r^a,r^b\}$ bifurcate from the zero value at the
first transition and this emergence of non-factorizability
stabilizes here the intermediate PVB-AF phase (compare Figs.
\ref{diagns} and \ref{diag}) and persists in the $G$-AF phase
where they overlap again ($r^a=r^b$). In the regime of PVB-AF
phase we observe also almost linear decrease of the in-plane
$R^{a,b}$ staying close to each other and a smaller drop of $R^c$.
Although these quantities are all small, the order parameters (see
Fig. \ref{w2}) are small too, so we conclude that spin-orbital
entanglement is qualitatively important here. The minimum of all
$R^{\gamma}$ is located at the second transition indicating that
highly entangled states play a role also at the onset of the
$G$-AF phase.

\begin{figure}[t!]
    \includegraphics[width=8.2cm]{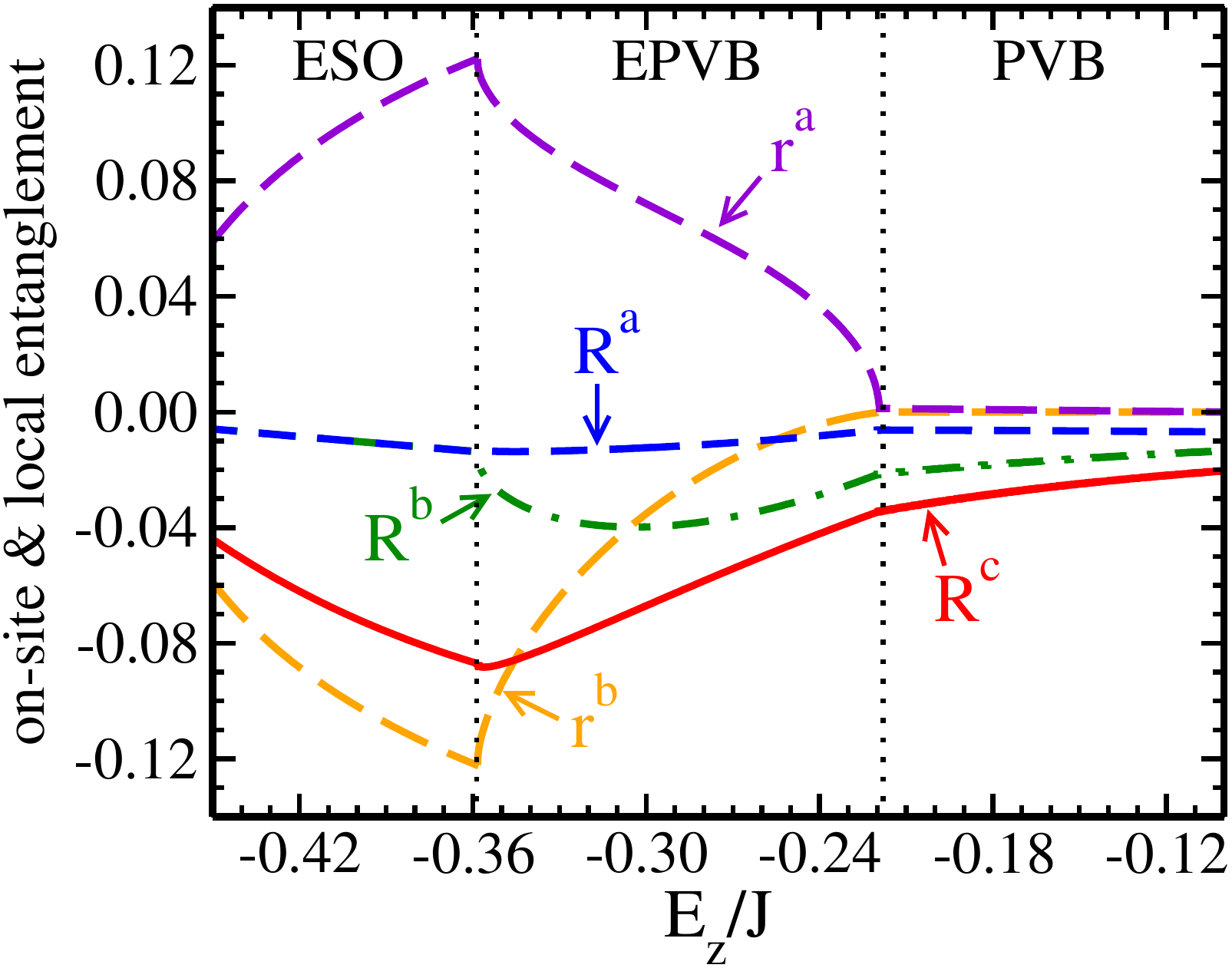}
    \caption{(Color online) On-site $r^{\gamma}$ Eq. (\ref{entalo})
and bond $R^{\gamma}$ Eq. (\ref{entabo}) entanglement parameters
for $\eta =0.15$ and $-0.45<E_z/J<-0.1$ in the
                  ESO, EPVB and PVB phases.}
    \label{e3}
\end{figure}

Figure \ref{e3} shows spin-orbital entanglement in the most exotic
part of the phase diagram with the ESO, EPVB and PVB phases for
$\eta =0.15$ and $-0.45<E_z/J<-0.1$. The on-site spin-orbital
covariances $\{r^a,r^b\}$ take high, opposite values in both the
ESO and EPVB phase, with maximum (minimum) at the transition line
between them. Comparing to other phases $r^{a,b}$ values are
highest in the ESO and EPVB phases, and comparing the two phase
diagrams in Figs. \ref{diagns} and \ref{diag}, we recognize that
spin-orbital entanglement is a constitutive feature of both ESO
and EPVB states. We emphasize that the on-site spin-orbital
entanglement is strong and complementary in the ESO phase on the
bonds along the $a$ and $b$ direction ($r^a=-r^b$), while it
vanishes between the $ab$ planes ($r^c=0$). These results indicate
spin-orbital fluctuations in the $ab$ planes, with $\langle
S^z\sigma^x\rangle\neq 0$ and no fluctuations along the $c$ axis,
where $r^c$ follows from $\langle S^z\sigma^z\rangle=0$. In
contrast, in the EPVB phase there is also finite on-site
entanglement for the interlayer order parameters, $r^c\neq 0$.

\begin{figure}[t!]
    \includegraphics[width=8.2cm]{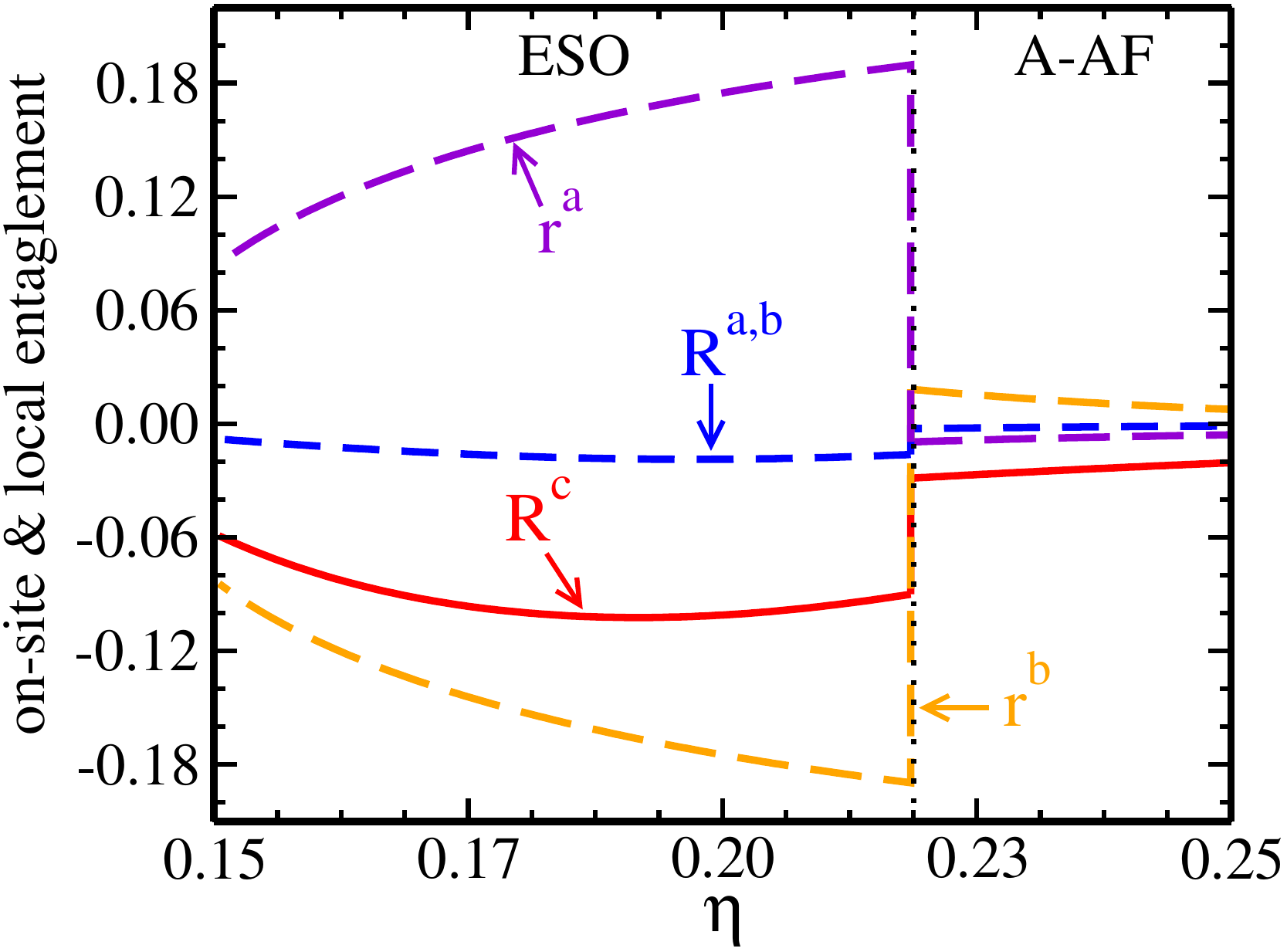}
    \caption{(Color online) On-site $r^{\gamma}$ Eq. (\ref{entalo})
and bond $R^{\gamma}$ Eq. (\ref{entabo}) entanglement parameters
for $E_z=-0.43J$ and $0.15<\eta<0.25$ in the
                  ESO and $A$-AF phases.}
    \label{e4}
\end{figure}

Looking at the bond parameters $R^{\gamma}$ we see that the
dominant one is $R^c$ falling gradually in the ESO down to the
minimum at the ESO-EPVB transition. At the same point $R^{b}$
drops from zero value in the ESO phase and takes maximally
negative value inside the EPVB regime. In contrast, $R^{a}$
remains close to zero in the entire regime of parameters and in
the PVB all the covariances go to zero showing that the order
within the PVB phase is practically disentangled. The dominant
role of $R^c$ comes from the $c$--axial symmetry of the ESO phase
and increased quantum fluctuations on the ESO-EPVB border while
the non-zero value of $R^{b}$ in the EPVB phase follows from the
magnetic and orbital order on the cube in the $bc$ plane mentioned
in the previous section.

When Hund's exchange is increased across the transition between
the ESO and $A$-AF phases, one finds that bond and on-site
spin-orbital covariances are radically different in both phases,
see Fig. \ref{e4}. The plot shows that the ESO phases is much
stronger entangled than the $A$-AF one where all the covariances
stay close to zero. Only the in-plane $R^{a,b}$ parameters are
small also in the ESO phase but the other covariances, including
the bond covariance along the $c$ axis $R^c$, take considerable
values.

\begin{figure}[t!]
    \includegraphics[width=8.2cm]{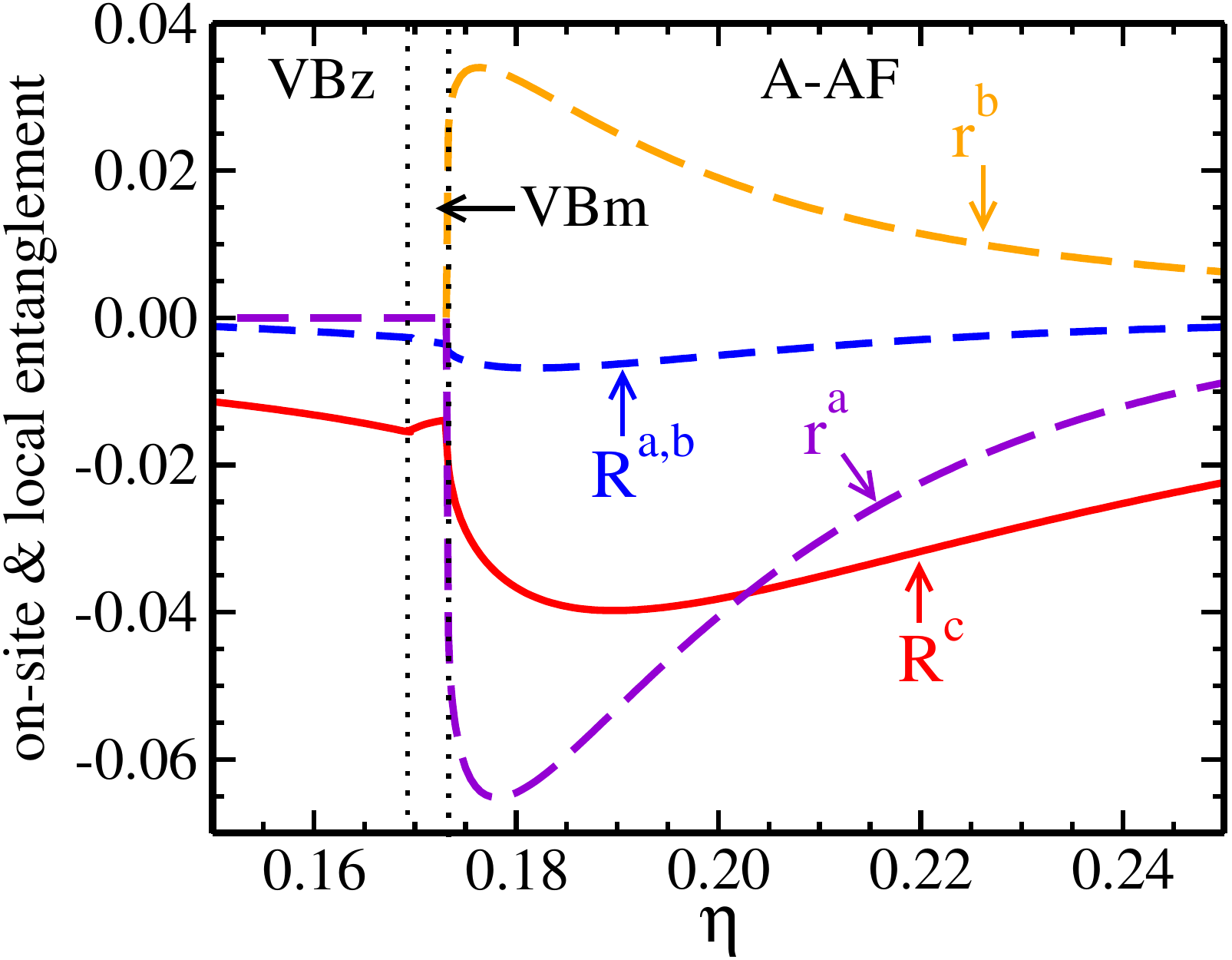}
    \caption{(Color online) On-site $r^{\gamma}$ Eq. (\ref{entalo})
and bond $R^{\gamma}$ Eq. (\ref{entabo}) entanglement parameters
for $E_z=-0.72J$ and $0.15<\eta<0.25$ in the
                  VB$z$, VBm and $A$-AF phases.}
    \label{e5}
\end{figure}

Finally, we focus on the range of large negative crystal field
splitting $E_z=-0.72J$ and display the spin-orbital covariances in
the VB$z$, VBm and $A$-AF phases for increasing Hund's exchange
$0.15<\eta<0.25$, see Fig. \ref{e5}. On the one hand, looking at
the VBm region of the plot we can understand why this phase can
exist when factorized spin-orbital MF is applied (again, compare
Figs. \ref{diagns} and \ref{diag}); the on-site covariances
$\{r^a,r^b\}$ vanish here and within the VB$z$ phase. On the other
hand, one finds certain on-site entanglement in the $A$-AF phase,
especially close to the transition line --- this shows why the
$A$-AF area is expanded in Fig. \ref{diag} as compared with the
non-factorized phase diagram of Fig. \ref{diagns}. Concerning bond
entanglement, it is significant (finite $R^c<0$) only along the
interlayer $c$ bonds in all these three phases, taking maximal
values of $|R^c|$ in the $A$-AF phase. One can understand this as
follows: in the VB$z$ phase the orbital order is almost perfect
and orbitals stay frozen --- therefore spin-orbital factorization
is here almost exact as indicated by a low value of $R^c$. This is
not the case in the $A$-AF phase where orbitals fluctuate,
especially close to the transition line to the VBm phase. The
$R^{a,b}$ bond parameters are small due to the imposed FM order
within the $ab$ planes which decouples the spin from orbital
fluctuations on the bonds along the $a$ and $b$ directions.

\section{Discussion and summary}
\label{sec:summa}

The numerical results presented in the last three sections,
obtained using the sophisticated mean-field approach with an
embedded cubic cluster, provide a transparent and rather complete
picture of possible ordered and disordered phases in the bilayer
spin--orbital $d^9$ model. This approach is well designed to
determine the character of spin, orbital and spin-orbital order
and correlations in all the discussed phases as it includes the
most important quantum fluctuations on the bonds and captures the
essential features of spin-orbital entanglement. By analyzing
order parameters we also presented evidence which allowed us to
identify essential features of different phases and to distinguish
between first and second order phase transitions. This is
especially important in cases when two phases are separated by an
intermediate configuration, such as the PVB-AF or VBm phase, where
one finds a gradual evolution from singlet to AF correlations,
supported (or not) by non-factorizable spin-orbital order
parameter. We believe that the cluster mean-field approach
presented here and including thie joint spin-orbital order
parameter is more realistic because there is no physical reason,
apart from the form of the $d^9$ Hamiltonian, to treat spin and
orbital operators as the only fundamental symmetry-breaking
degrees of freedom in any phase.

Interestingly, the results show that the bottom part of the phase
diagram of the $d^9$ spin-orbital model does not exhibit any
frustration or spin-orbital entanglement up to $\eta\simeq 0.07$
and the type of spin and orbital order are chosen there
predominantly by the crystal field term $\propto E_z$. Quantum
fluctuations dominate at $E_z\simeq 0$ and for $E_z<0$, where they
stabilize either in-plane or interplanar spin singlets accompanied
by directional orbitals which stabilize them. These two phases are
reminiscent of the resonating valence-bond (RVB) phase and PVB
phase found in the 3D spin-orbital $d^9$ model.\cite{Fei97} Here,
however, the VB$z$ phase extends down to large values of $E_z$,
where instead the long-range order in the $G$-AF$z$ phase was
found in the 3D model. This demonstrates that interlayer quantum
fluctuations are particularly strong in the present bilayer case.
On the contrary, at $E_z>0$ one finds the $G$-AF spin order which
coexists with FO order of $x$ orbitals. It is clear that both the
VB$z$ and $G$-AF phase are favored by the interplay of lattice
geometry and by the shape of occupied orbitals for
$E_z\to\pm\infty$. In this low-$\eta$ regime of the diagram the
area occupied by the PVB phase is narrow and especially orbital
order is affected by the quantum critical fluctuations. The planar
singlets are formed shortly after leaving the VB$z$ phase and
remain stable afterwards. Spin-orbital non-factorizability seems
to be marginal in the entire VB$z$ phase but plays certain role
when switching to the planar singlet phase, especially visible for
the interplane bond covariance $R^c$.

On the contrary, in the PVB phase away from the critical regime
spin-orbital non-factorizability vanishes and suddenly reappears in the
$G$-AF phase, not as a transition effect but rather as a robust feature
vanishing only for high values of $E_z$. We argue that this is related
with surprisingly rigid interplane spin-spin correlations which should,
if we think in spin-orbital factorizable way, decay quickly as $t^c$
approaches $1/2$. Following this "factorizable reasoning" we
could also expect stability of the $C$-AF phase for higher values of
$\eta$, above the $G$-AF phase. These effects are absent in our results,
showing that intuition suggesting spin-orbital factorization can be
misleading even when considering such a simple isotropic orbital
configuration.

For higher values of Hund's exchange $\eta$ frustration increases
when AF exchange interactions compete with FM ones, and as a
result the most exotic phases with explicit on-site spin-orbital
entanglement arise; two of them, the ESO and EPVB phase are
neighboring and placed in between the VB$z$ and PVB ones, and they
become degenerate with both of them at the multicritical point
where four phases meet (see Fig. \ref{diag}). This situation
follows from the fact that singlet phases are more susceptible to
ferromagnetism favored by high $\eta$ than the $G$-AF phase is,
which turned out to be surprisingly robust. The ESO phase is also
a singlet phase similar to the VB$z$ one but with spin singlets
and orbital order gradually suppressed under increasing Hund's
exchange $\eta$. At the same time spin-orbital order stays almost
constant and spin-orbital entanglement grows.

Further increase of $\eta$ always leads to the $A$-AF phase
throughout a discontinuous transition accompanied by an abrupt
drop of spin-orbital entanglement. Above $\eta\simeq 0.2$ the ESO
phase is completely immersed in the $A$-AF one and ends up with a
single bicritical point. If we come back below $\eta\approx 0.2$
then the ESO changes smoothly into the EPVB phase, being an
entangled precursor of the PVB order, meaning that the non-uniform
orbital order and in-plane singlets are formed and spin-orbital
entanglement drops. On the other hand, this phase can be also seen
as an extension of the $A$-AF into fully AF sector because the
EPVB phase has long-range magnetic order, being however strongly
non-uniform (see Fig. \ref{c3}).

Finally, we would like to remark that experimental phase diagrams
of strongly correlated transition metal oxides are one of the
challenging directions of recent research. Systematic trends
observed for the onset of the magnetic and orbital order in the
$R$VO$_3$ perovskites have been successfully explained by the
competing interactions in presence of spin-orbital
entanglement.\cite{Hor08} In contrast, the theory could not
explain exceptionally detailed information on the phase diagram of
the $R$MnO$_3$ manganites which accumulated due to impressive
experimental work.\cite{Goo06} The present K$_3$Cu$_2$F$_7$
bilayer system is somewhat similar to
La$_{2-2x}$Sr$_{1+2x}$Mn$_2$O$_7$ bilayer manganites with very
rich phase diagrams and competition between phases with different
types of long-range order in doped systems.\cite{bilay} Such
phases are generic in transition metal oxides and were also
reproduced in models of bilayer manganites which have to include
in addition superexchange interaction between core $t_{2g}$
spins\cite{Ole03,Dag06} that suppresses spin-orbital fluctuations
and entanglement in the $e_g$ subsystem. In contrast,
K$_3$Cu$_2$F$_7$ bilayer is rather unique as the only electronic
interactions arise here due to entangled spin-orbital
superexchange. They explain the origin of the VB$z$ phase
observed\cite{Man07} in K$_3$Cu$_2$F$_7$ but not found in bilayer
manganites, and provide a unique opportunity of investigating
whether signatures of spin-orbital entanglement could be
identified in future experiments.

Summarizing, the presented analysis demonstrates that spin-orbital
entanglement plays a crucial role in complete understanding of the
phase diagram of the bilayer spin-orbital $d^9$ model. By
introducing additional spin-orbital order parameter independent of
spin and orbital mean fields we obtained phases with spin disorder
in highly frustrated regime of parameters. The example of the
entangled ESO and EPVB phases shows that joint spin-orbital order
can be at least as strong as the other two (spin or orbital) types
of order, or may even persist as the only symmetry breaking field
when the remaining ones vanish. We argue that the cluster method
we used here is sufficiently realistic to investigate the phase
diagram of the 3D spin-orbital $d^9$ model, and could be applied
to other spin-orbital superexchange models adequate for undoped
transition metal oxides.

\acknowledgments

We thank Joachim Deisenhofer, Lou-Fe' Feiner and Krzysztof
Ro\'sciszewski for insightful discussions.
We acknowledge support by the Foundation for Polish Science (FNP)
and by the Polish Ministry of Science and Higher Education under
Project No. N202 069639.

\appendix*

\section{Solution of the mean-field equations}
\label{sec:app}

Here we present briefly the solution of the self-consistency Eqs.
(\ref{SCeq}) and (\ref{SCeq1}) obtained in the single-site MF
approximation. It is obtained as follows: assuming $a$--$b$
symmetry of the system, i.e., putting $\chi^a=\chi^b$ and
$\xi^a=\xi^b$, we derive $t^a$ and $t^c$ from Eq. (\ref{abdef}) as
functions of $\alpha$ and $\beta$,
\begin{eqnarray}
\label{tatc}
t^c&=&4g\alpha +2g E_z +g_1,  \\
t^a&=&-2g\alpha -\frac{2\beta}{\sqrt{3}}g_2 -\frac{1}{2}(2gE_z+g_1),
\end{eqnarray}
with
\begin{eqnarray}
g&=&(\chi^a-\xi^a+2\chi^c-2\xi^c)^{-1},   \\
g_1&=&g(\xi^a-\xi^c),   \\
g_2&=&(\chi^a-\xi^a)^{-1}.
\end{eqnarray}
Now we introduce a parametrization
\begin{equation}
\alpha =\Delta\sin\phi, \hskip 1cm \beta=\Delta\cos\phi,
\end{equation}
and use the self-consistency Eqs. (\ref{SCeq}) and (\ref{SCeq1}).
From $t^c$ one finds immediately $\sin{\phi}$ depending on
$\Delta$,
\begin{equation}
\sin{\phi }=-2\,\frac{2gE_z+g_1}{8g\Delta +1}.
\label{sin}
\end{equation}
Comparing Eqs. (\ref{SCeq}) and (\ref{tatc}) for $t^a$ one gets:
\begin{eqnarray}
&&\sin\phi-\sqrt{3}\cos{\phi}=\nonumber\\
&-&4\left(2g\Delta \sin{\phi}+\frac{2}{\sqrt{3}}g_2\Delta\cos{\phi}
+gE_z+\frac{1}{2}g_1\right).
\label{selfco}
\end{eqnarray}
After inserting $\sin{\phi}$ into Eq. (\ref{selfco}) we obtain
a surprisingly simple result for $\cos{\phi}$:
\begin{equation}
\cos{\phi }\left(\Delta-\frac{3}{8g_2}\right) =0.
\label{cos}
\end{equation}
This leads to two classes of solutions of self-consistency Eqs.
(\ref{SCeq}) and (\ref{SCeq1}): (i) either $\cos{\phi}=0$, or (ii)
$\Delta=3/8g_2$ and $\cos{\phi}\not=0$. The first option implies
$\sin{\phi}=\pm 1$ and leads to two uniform orbital configurations
with $t^c=\mp \frac{1}{2}$ and $t^a=t^b=-t^c/2$. Furthermore,
using Eq. (\ref{sin}) we can calculate $\Delta$ and find the
borders of these uniform phases demanding $\Delta\geq 0$.

The second option, i.e., $\cos{\phi}\not=0$, implies
AO-type of order with:
\begin{eqnarray}
t_c\!\!&=&\!\frac{2gE_z+g_1}{3g/g_2 +1},\\
t_a\!\!&=&\!-\frac{1}{2}\frac{2gE_z+g_1}{3g/g_2 +1}
\mp\frac{\sqrt{3}}{2}\sqrt{\frac{1}{4}
-\left(\frac{2gE_z+g_1}{3g/g_2 +1}\right)^2},
\end{eqnarray}
and with phase borders defined by the condition:
$2|t^c|\leq 1$. The phase borders given here set the maximal range of
the phase under consideration and cannot be treated as the lines of
phase transitions shown in the phase diagram; the latter lines are
determined by comparing the ground state energies $E_0$ calculated
form Eq. (\ref{Emin}).


\begin{thebibliography}{99}
\bibitem{noteri} For convenience, we use here the same notation
                 for the multiplet coefficients $r_i=\{r_1,r_2,r_4\}$
                 as in Ref. \onlinecite{Ole05}.
\end{thebibliography}
\end{document}